\documentclass[aps,prb,twocolumn,superscriptaddress,eqsecnum,showpacs,showkeys,amsmath,amssymb]{revtex4}

\usepackage{amsfonts}
\usepackage{amssymb,amsmath}
\usepackage{graphicx}
\usepackage{dcolumn}
\usepackage{bm}
\usepackage{color}

\begin{document}

\title{Emergent Ising degrees of freedom in frustrated two-leg ladder and bilayer $s=1/2$ Heisenberg antiferromagnets}

\author{Oleg Derzhko}
\affiliation{Institute for Condensed Matter Physics,
          National Academy of Sciences of Ukraine,
          1 Svientsitskii Street, L'viv-11, 79011, Ukraine}
\affiliation{Institut f\"{u}r theoretische Physik,
          Universit\"{a}t Magdeburg,
          P.O. Box 4120, 39016 Magdeburg, Germany}

\author{Taras Krokhmalskii}
\affiliation{Institute for Condensed Matter Physics,
          National Academy of Sciences of Ukraine,
          1 Svientsitskii Street, L'viv-11, 79011, Ukraine}

\author{Johannes Richter}
\affiliation{Institut f\"{u}r theoretische Physik,
          Universit\"{a}t Magdeburg,
          P.O. Box 4120, 39016 Magdeburg, Germany}

\date{\today}



\begin{abstract}
Based on exact diagonalization data
for finite quantum Heisenberg antiferromagnets on two frustrated lattices (two-leg ladder and bilayer)
and analytical arguments
we map low-energy degrees of freedom of the spin models in a magnetic field on classical lattice-gas models.
Further we use transfer-matrix calculations and classical Monte Carlo simulations
to give a quantitative description of low-temperature thermodynamics of the quantum spin models.
The classical lattice-gas model yields an excellent description 
of the quantum spin models up to quite large temperatures.
The main peculiarity of the considered frustrated bilayer
is a phase transition
which occurs at low temperatures for a wide range of magnetic fields below the saturation magnetic field
and belongs to the two-dimensional Ising model universality class.
\end{abstract}

\maketitle

\section{Introduction}
\label{sec1}

Antiferromagnetically interacting Heisenberg spins on geometrically frustrated lattices
have attracted much attention of physicists during last years.\cite{lhuillier,mikeska,richter_lnp}
A rapidly developing direction in this area
is the study of the properties of such models in the presence of an external magnetic field.
The recent finding
that a wide class of geometrically frustrated quantum spin antiferromagnets
(including kagom\'{e}, checkerboard and pyrochlore lattices)
has quite simple ground states in the vicinity of the saturation field
-- the so-called independent localized-magnon states\cite{loc_mag,loc_mag_review} --
has further stimulated studies of the corresponding frustrated quantum antiferromagnets
at high magnetic fields.
Among others,
we mention here the recent papers
concerning the detailed analysis
of the low-temperature high-field magnetothermodynamics
of a number of one-, two-, and even three-dimensional frustrated quantum antiferromagnets
which support localized-magnon states.\cite{zhi1,zhi2,SP,zhi3,zhi4,d&r_a,d&r_b,d&r_c,d&r_d}
Thus,
the low-energy degrees of freedom
of the quantum Heisenberg antiferromagnet on a kagom\'{e} lattice
in the vicinity of the saturation field
can be mapped onto a gas of hard hexagons on a triangular
lattice.{\cite{zhi1,d&r_a,zhi3}
The latter model exhibits a phase transition\cite{baxter}
that implies a phase transition in the spin model
at finite (low) temperatures below (but close to) the saturation field.
Similarly,
the low-energy degrees of freedom of the checkerboard antiferromagnet
can be mapped onto a gas of hard squares on a square lattice\cite{zhi4,d&r_a}
with the size of squares
which corresponds to nearest-neighbor and next-nearest-neighbor exclusion
and the resulting lattice-gas model also exhibits a phase transition\cite{large_hs}
that implies the corresponding peculiarities of the spin model
at low temperatures below the saturation field.
Although the performed analysis\cite{zhi1,zhi3,zhi4}
suggests interesting examples of the two-dimensional Heisenberg system
with a phase transition at high magnetic fields and low temperatures,
the results for the kagom\'{e} and checkerboard antiferromagnets
cannot be considered as conclusive examples,
since not all of the relevant low-energy states
are included in the hard-core-object lattice-gas
description.\cite{zhi1,zhi3,zhi4,d&r_a,d&r_b,d&r_c,d&r_d,linear_ind}
The effect of these additional states  on the thermodynamic properties for
both models remains an unresolved problem.
In a recent paper we have discussed another two-dimensional frustrated quantum Heisenberg antiferromagnet
-- a frustrated bilayer.
The low-energy degrees of freedom of that model around the saturation field
can be mapped on a hard-square model
(hard squares on a square lattice corresponding to nearest-neighbor exclusion only),
see Refs.~\onlinecite{d&r_c} and \onlinecite{d&r_d}.
Contrary to the kagom\'{e} and checkerboard antiferromagnets,
for the frustrated bilayer antiferromagnet the hard-square states
completely exhaust all low-energy states of the spin model
in the vicinity of the saturation field
and all other low-lying excited states are separated by a finite energy gap.
Therefore a phase transition inherent in the hard-square
model\cite{baxter_hs}
leads to a phase transition for the spin model at high magnetic fields and low temperatures
which cannot be questioned.
We also note here that in spite of the fact that the Mermin-Wagner theorem forbids
long-range order for the two-dimensional Heisenberg model
at any non-zero temperature at zero field,\cite{mermin}
in the presence of an external magnetic field it may be indeed present.
We will show in our paper 
that emergent discrete degrees of freedom 
may lead to Ising-like antiferromagnetic long-range order at low temperatures.

In the present paper,
we extend substantially the studies of low-temperature properties
for the frustrated bilayer
reported in Refs.~\onlinecite{d&r_c} and \onlinecite{d&r_d}
and the frustrated two-leg ladder
reported in Refs.~\onlinecite{ho_mi_tr}, \onlinecite{d&r_b}, and \onlinecite{d&r_d}
now taking into account within lattice-gas description not only the highly
degenerate ground-state manifold but also low-lying excited states.
As a result we arrive at a lattice-gas model with finite nearest-neighbor
repulsion.
For that effective model we perform
the transfer-matrix calculations (for the frustrated ladder)
and classical Monte Carlo simulations (for the frustrated bilayer)
to examine the low-temperature behavior of the quantum spin model for a wide region of the magnetic field.
Moreover, due to the inclusion of the excited
states the lattice-gas description excellently describes  the spin physics up to significantly higher
temperatures and in a much wider range of the magnetic field
in comparison with earlier studies.\cite{d&r_a,d&r_b,d&r_c,d&r_d}
The main message of our study is as follows:
Geometrical frustrations may lead to a simple structure of low-lying energy levels
which can be mapped onto a classical lattice-gas model
and, as a result, transfer-matrix calculations or classical Monte Carlo simulations provide a very good
description  of the low-temperature physics of the quantum spin model.
The most prominent result concerns the phase transition in the two-dimensional case
which ``survives'' if low-lying excited states are taken into account.

The theoretical investigation of the quantum Heisenberg antiferromagnet
on the two-leg ladder and square-lattice bilayer
has attracted a lot of attention during last years.
So far the main focus was on ground-state properties,
see, e.g.,
Refs.~\onlinecite{gelfand,lit_ladder,mila,ho_mi_tr,fouet,chandra,nedko} for the ladder
and Ref.~\onlinecite{bilayer} for the bilayer.
In our study we concentrate on low-temperature properties of these models.
We also note that the models under considerations are known as models with local conservation laws,
see Refs.~\onlinecite{gelfand,mila,ho_mi_tr,fouet,chandra,nedko,lc_dipla,HJS}.
On the other hand, these models belong to a class of systems which support localized-magnon states.\cite{d&r_a,d&r_b,d&r_c,d&r_d}

The paper is organized as follows.
In Sec.~\ref{sec2}
we introduce the quantum spin models and discuss their symmetries.
In Sec.~\ref{sec3}
we briefly present a spectroscopic study of the spin models.
In particular, we focus on a class of simple product eigenstates
[independent (isolated) localized magnons and interacting (overlapping) localized magnons]
which become the low-energy states under certain conditions.
In Sec.~\ref{sec4}
we map the localized-magnon states on the lattice-gas-model states
and discuss the degeneracy of the localized-magnon states.
In Sec.~\ref{sec5}
we calculate the contribution of the independent localized-magnon states 
to the thermodynamic quantities,
whereas
in Sec.~\ref{sec6}
we extend calculation of thermodynamic quantities taking into account the set of interacting localized-magnon states in addition.
In these sections
we compare lattice-gas model results with exact diagonalization data
for finite spin systems
to find the region of validity for the lattice-gas-model description.
Moreover,
we obtain the low-temperature thermodynamic quantities of the quantum spin models on the basis of
(i) transfer-matrix calculations for the one-dimensional case
and
(ii) classical Monte Carlo simulations for large two-dimensional lattice-gas systems.
A brief summary 
is given in the last section (Sec.~\ref{sec7}).
Some lengthy formulas for one- and two-dimensional lattice-gas models are collected in two appendices.

\section{Model}
\label{sec2}

We consider the antiferromagnetic Heisenberg model of $N=2{\cal{N}}$ quantum spins $s=1/2$
on the two lattices shown in Fig.~\ref{fig1}.
\begin{figure}
\begin{center}
\includegraphics[clip=on,width=7.5cm,angle=0]{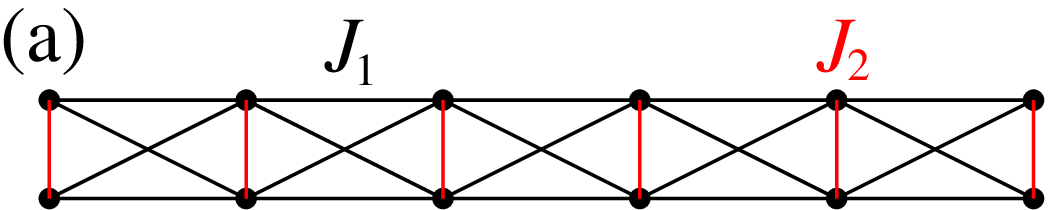}\\
\vspace{5mm}
\includegraphics[clip=on,width=7.5cm,angle=0]{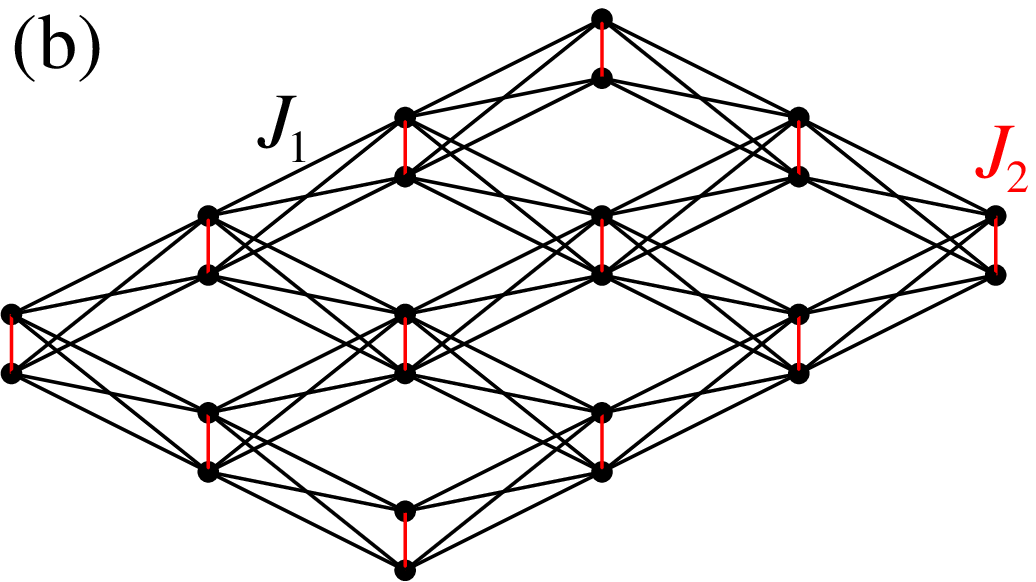}
\caption
{(Color online)
Lattices considered in this paper:
(a) the frustrated two-leg ladder
and
(b) the frustrated bilayer.
The vertical bonds have the strength $J_2>0$
whereas all other bonds have the strength $J_1>0$.}
\label{fig1}
\end{center}
\end{figure}
The Hamiltonian of the model reads
\begin{eqnarray}
\label{2.01}
H=\sum_{(pq)}J_{pq}{\bf{s}}_p\cdot{\bf{s}}_q-hS^z.
\end{eqnarray}
Here
the sum runs over the bonds
which connect the neighboring sites
on the spin lattice shown in Fig.~\ref{fig1},
$J_{pq}>0$ are the antiferromagnetic exchange constants
between the sites $p$ and $q$
which take two values,
namely,
$J_2$ for the vertical bonds
and
$J_1$ for all other bonds,
$h\ge 0$ is the external magnetic field,
and
$S^z=\sum_ps_p^z$ is the $z$-component of the total spin.
In our study we imply periodic boundary conditions.
Further we set $J_1=1$ if not stated otherwise explicitly.

We introduce an underlying lattice of ${\cal{N}}=N/2$ sites.\cite{footnote1}
For the frustrated two-leg ladder that is a simple chain
whereas for the frustrated bilayer it is a square lattice.
Now it is convenient to denote the spin lattice sites as $m,i$,
where $m$ numbers the vertical bonds, 
i.e., it runs over all sites of the underlying lattice
and the index $i$ refers either to the lower ($i=1$) or to the upper ($i=2$) leg or layer.
Note that the underlying lattices are bipartite ones,
i.e., we can divide it into two sublattices $A$ and $B$,
and any two neighboring sites on the lattice belong to different sublattices.
Introducing further
the total spin of a vertical bond
${\bf{t}}_m={\bf{s}}_{m,1}+{\bf{s}}_{m,2}$
the spin Hamiltonian (\ref{2.01}) can be rewritten as follows
\begin{eqnarray}
\label{2.02}
H=
\sum_{m}\left[\frac{J_2}{2}\left({\bf{t}}_{m}^2-\frac{3}{2}\right)-ht^z_{m}\right]
+J_1\sum_{(mn)}{\bf{t}}_{m}\cdot{\bf{t}}_{n}.
\end{eqnarray}
Here the first sum runs over all sites of the underlying lattice
and the second sum runs over all bonds
which connect the neighboring sites on the underlying lattice.

The Hamiltonian (\ref{2.02}) depends
on the total spin of each vertical bond ${\bf{t}}_{m}$, $m=1,\ldots,{\cal{N}}$ only,
and the value of the total spin of a vertical bond is a good quantum number.
As a consequence,
the properties of the considered models can be studied in much more detail.
In particular, a large number of eigenstates can be constructed exactly, see
Refs.~\onlinecite{ho_mi_tr,d&r_b,d&r_c,d&r_d} and Sec.~\ref{sec3}.
In Sec.~\ref{sec3} we give a precise description of low-energy eigenstates of the spin Hamiltonian
in a regime when $J_2/J_1$ is sufficiently large (strong-coupling regime).
We again emphasize
that our focus is the low-temperature thermodynamics of these models in the strong-coupling regime
and therefore we will be interested not only in the energies of low-energy eigenstates
but also in their degeneracies, see Sec.~\ref{sec4}.

\section{Product eigenstates}
\label{sec3}

In this section we briefly summarize some known facts on a class of simple
product eigenstates of the spin Hamiltonian (\ref{2.01}) [or (\ref{2.02})]
which become the low-energy ones under certain conditions.\cite{ho_mi_tr,d&r_b,d&r_c,d&r_d}
For this purpose we may consider the subspaces with different values of $S^z$ separately
since the Hamiltonian (\ref{2.01}) commutes with the operator $S^z$.
We may assume at first $h=0$ since adding of the Zeeman contribution is trivial.
Obviously, the fully polarized state
$\vert\uparrow,\ldots,\uparrow\rangle$
is an eigenstate of the Hamiltonian (\ref{2.01}) with $S^z=N/2={\cal N}$.
The energy of this state is
$E_{\rm{FM}}={\cal{N}}J_1+{\cal{N}}J_2/4$
or
$E_{\rm{FM}}=2{\cal{N}}J_1+{\cal{N}}J_2/4$
for the one- or two-dimensional case, respectively.
This state is the ground state for high magnetic fields.

Next we consider eigenstates
\begin{eqnarray}
|n\rangle=|0_{m_1}\rangle|0_{m_2}\rangle\cdots|0_{m_n}\rangle |{\rm{FM}}\rangle_{{\rm{R}}} ,
\nonumber \\
|0_{m_i}\rangle=\frac{1}{\sqrt{2}}\left(\vert \uparrow_{m_i,1}\downarrow_{m_i,2}\rangle
-\vert \downarrow_{m_i,1}\uparrow_{m_i,2}\rangle\right),
\label{3.01}
\end{eqnarray}
where a subset of $n$, $1 \le n \le {\cal N}$ spin pairs on vertical bonds
$m_1$, $m_2$, \ldots, $m_n$ are in a singlet state $|0_{m}\rangle$
and the remaining ${\cal N} - n$ other spin pairs on vertical bonds are in a fully polarized triplet state
$\vert \uparrow_{t,1}\uparrow_{t,2}\rangle$  
with ${\bf{t}}^2_t=2$ and $t^z_t=1$, 
where the index $t$ labels the vertical bond carrying the triplet.
The triplet bonds form a fully polarized ferromagnetic background $|{\rm{FM}}\rangle_{\rm{R}}$.
Obviously,  these states have a magnetization $S^z={\cal N} - n$.
Each vertical singlet contributes with $-3J_2/4$ to the energy.
The contribution of a polarized triplet at a vertical bond $t$ is $J_2/4 + \gamma_t J_1$,
where $\gamma_t$ counts the number of neighboring triplets of the triplet at
certain vertical bond $t$.
Hence, for the energy of the state (\ref{3.01}) 
we get $E_n=-3nJ_2/4 + ({\cal{N}}-n)J_2/4 + J_1\sum'_{t}\gamma_t/2$.
It remains to calculate $\sum'_{t}\gamma_t$, where the sum runs over all ${\cal{N}}-n$ vertical triplet bonds.
To get a state of minimal energy in a certain sector of $S^z$ we have to minimize $\sum'_{t}\gamma_t$.

Obviously,
we get the minimal $\sum'_{t}\gamma_t=2({\cal{N}}-2n)$ in the one-dimensional case
or $\sum'_{t}\gamma_t=4({\cal{N}}-2n)$ in the two-dimensional case
in all sectors ${\cal N}/2 \le S^z < {\cal N}$,
if we have no neighboring singlets (hard-core rule).
This is a weak  constraint, and,
consequently, there are many states fulfilling this constraint.
Note that these states belong to the
class of so-called independent localized-magnon states appearing in many frustrated
lattices.\cite{richter_lnp,loc_mag,loc_mag_review,zhi2,zhi1,zhi3,zhi4,d&r_a,SP,d&r_b,d&r_c,d&r_d}
The energy of these localized-magnon states is
\begin{equation}
E^{\rm{lm}}_n=E_{{\rm{FM}}}-n\epsilon_1 ,
\label{3.02}
\end{equation}
where
$\epsilon_1 =J_2+2J_1$
or
$\epsilon_1=J_2+4J_1$
for the ladder or bilayer case, respectively.
The maximal number of independent localized magnons is $n_{\max}={\cal N}/2$.
For $n=n_{\max}$ there are two degenerate  localized-magnon states
where one sublattice ($A$ or $B$) carries the vertical singlets and the other one is occupied by vertical triplets
(so-called ``magnon-crystal'' states).
The energy of a magnon-crystal state is $E_{{\cal N}/2}^{\rm{lm}}=-{\cal N}J_2/4$.
In Ref.~\onlinecite{d&r_b} and \onlinecite{d&r_c} (see also Ref.~\onlinecite{schmidt})
it was shown that the independent localized-magnon states are ground states
in the respective sectors of $S^z$ if $J_2 \ge 2J_1$ for the ladder and  $J_2 \ge 4J_1$ for the bilayer.
In a magnetic field we have the energy 
$E_n^{\rm{lm}}(h)=E_n^{\rm{lm}}-hS^z=E_{{\rm{FM}}}-n\epsilon_1-h(N/2 - n)$.
Hence all these independent localized-magnon states are degenerate at $h=h_1=\epsilon_1$,
where $h_1=\epsilon_1$ is the saturation field.
As a result one finds for the ground-state magnetization $M(T=0,h,N)$
the well-known jump to saturation at $h=h_1$ with a preceding wide plateau
(see, e.g., Refs.~\onlinecite{ho_mi_tr,richter_lnp,loc_mag}),
illustrated in Fig.~\ref{fig2}, 
where the plateau state is a two-fold degenerate magnon-crystal state.
\begin{figure}
\begin{center}
\includegraphics[clip=on,width=7.5cm,angle=0]{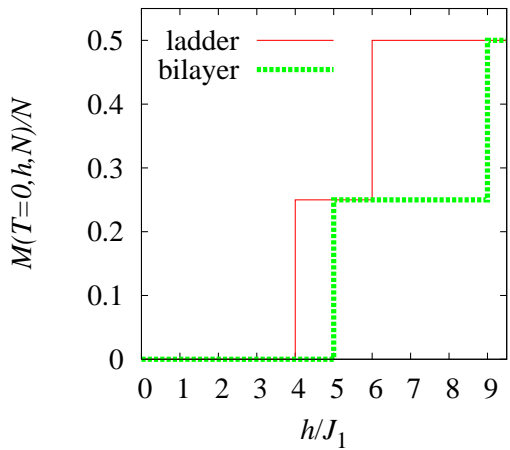}
\caption
{(Color online)  
Ground-state magnetization curves for the frustrated ladder (thin solid) and bilayer (thick broken)
in the considered strong-coupling regime.
We set $J_1=1$, $J_2=4$ (ladder) and $J_1=1$, $J_2=5$ (bilayer),
i.e., 
we have 
$h_1=6$ and $h_2=4$ for the ladder
and
$h_1=9$ and $h_2=5$ for the bilayer.}
\label{fig2}
\end{center}
\end{figure}

Product eigenstates with higher energies in the sectors ${\cal N}/2 \le S^z < {\cal N}-1$
are states where some of the $n$ vertical singlets are neighbors.
These states can be understood as non-independent (i.e., interacting) localized-magnon states.
Supposing that we have ${\nu}$ pairs of neighboring vertical singlets
then we get an energy
\begin{eqnarray}
E_n^{\nu}=E_n^{{\rm{lm}}}+\nu J_1,
\label{3.03}
\end{eqnarray}
where $J_1$ can be understood as the repulsion energy.
For large enough $J_2 > J_2^c$ eigenstates with $\nu=1$ are the lowest excitations
above the independent localized-magnon ground states for ${\cal N}/2 \le S^z < {\cal N}-1$.
Based on finite-size calculations ($N=32$)
we estimate $J_2^c \approx 3.00J_1$ for the ladder and $J_2^c \approx 4.65J_1$ for the bilayer.

Going to lower magnetization $0\le S^z < {\cal N}/2$ no independent localized-magnon states exist,
and  the ``interacting'' localized-magnon states with $n>{\cal N}/2$ vertical singlets can become ground states.
A lowest-energy state with $n={\cal N}/2 + r$, $r=1,\ldots,{\cal N}/2$ localized magnons
is, e.g., a state where ${\cal N}/2$ magnons (singlets on vertical bonds) occupy one sublattice completely
(i.e., they are in the ``magnon-crystal'' state)
and the remaining $r$ magnons sit on the other sublattice.
These states have a magnetization $S^z = {\cal N}/2-r$ and an energy
(now $\sum'_{t}\gamma_t=0$)
\begin{equation}
E_{\frac{{\cal{N}}}{2}+r}=-\frac{{\cal N}J_2}{4} - rJ_2.
\label{3.04}
\end{equation}
In a magnetic field we have the energy
$E_{{\cal{N}}/2+r}-hS^z=-{\cal N}J_2/4 - rJ_2 - h({\cal N}/2 - r)$.
Hence all these interacting localized-magnon states are degenerate at $h=h_2=J_2$.
As a result one finds another jump in the ground-state magnetization at $h=h_2$ with a 
preceding wide $S^z=0$ plateau,\cite{ho_mi_tr}
see Fig.~\ref{fig2},
where this plateau state is a non-degenerate state where all vertical bonds carry a singlet.
Low-lying excited states in the subspaces with $0\le S^z < {\cal N}/2$ are constructed
by rearranging the vertical singlets 
to increase the number of neighboring singlets to $\nu\ge 1$ 
(i.e., the  sublattice formerly completely occupied by singlets on vertical bonds 
is now incompletely occupied by singlets).
For these excited states the increase of energy of the resulting states again is given by $\nu J_1$,
see Eq. (\ref{3.03}).

\section{Degeneracy of localized-magnon states}
\label{sec4}

After having illustrated the basic facts on exact product eigenstates of the considered models
which become ground states and excited low-energy states in all subspaces with $S^z=N/2,\ldots,0$
for sufficiently strong vertical bonds $J_2>J_2^c$,
we will now calculate their degeneracies
using a mapping of low-energy degrees of freedom of the quantum spin model (\ref{2.01}) on appropriate classical lattice-gas models.
This kind of mapping was used for various frustrated lattices hosting independent localized-magnon
states.\cite{zhi1,zhi3,zhi4,d&r_a,d&r_b,d&r_c,d&r_d}
Let us mention here that recently it has been found that this kind of mapping is also applicable to some flat-band Hubbard models.\cite{hub}
Note that in all previous papers using such a mapping, 
it was restricted to the independent localized-magnon {\it ground states}, only.
Based on the spectroscopic analysis given in the previous section, 
here we overcome this restriction 
and extend the results for the ladder and the bilayer presented in Refs.~\onlinecite{d&r_b,d&r_c,d&r_d}
including the interacting localized-magnon {\it low-lying excited states}.

For a better understanding of the mapping of the interacting localized-magnon states 
that will be discussed in the next paragraph 
we start with a brief illustration of the mapping of the independent localized-magnon states,
see also Refs.~\onlinecite{d&r_b,d&r_c,d&r_d}.
As mentioned in the previous section for the independent localized-magnon states a hard-core rule is valid,
i.e., they cannot occupy neighboring sites on the underlying lattice 
(chain or square lattice).
Hence,
the number of  possibilities  to put $n$ independent localized magnons on the two-leg ladder (bilayer)
is equivalent to the number of possibilities to place $n$ hard dimers (hard squares) on a chain (square lattice) 
of ${\cal{N}}=N/2$ sites, cf. Fig.~\ref{fig3}(a).
\begin{figure}
\begin{center}
\includegraphics[clip=on,width=7.5cm,angle=0]{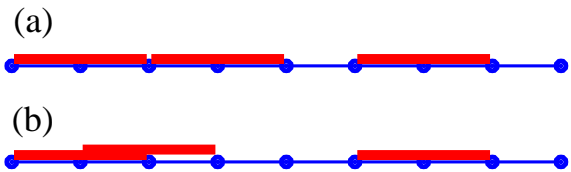}
\caption
{(Color online)
Hard-dimer description of the localized-magnon states of the frustrated two-leg ladder.
(a) Independent localized-magnon states (non-overlapping hard-dimer states).
(b) Interacting localized-magnon states (overlapping hard-dimer states).}
\label{fig3}
\end{center}
\end{figure}
Denoting the ground-state degeneracy in the $n$-magnon subspace by $g_{\cal{N}}(n)$,
we find for $n=0,1,\ldots,{\cal{N}}/2$ that $g_{\cal{N}}(n)={\cal{Z}}_{\rm{hc}}(n,{\cal{N}})$,
where ${\cal{Z}}_{\rm{hc}}(n,{\cal{N}})$ is simply the canonical partition function
of $n$ hard-core objects (hard dimers or hard squares)
on a ${\cal{N}}$-site lattice (chain or square lattice)
with periodic boundary conditions imposed.
We may call the independent localized-magnon states the hard-core states.
As mentioned above, the independent localized-magnon states
are ground states   in the subspaces with ${\cal{N}}/2\le S^z\le {\cal{N}}$
if $J_2>2J_1$ (ladder) or $J_2>4J_1$ (bilayer).
Moreover, 
they are linearly independent 
and form an orthogonal basis in each subspace with ${\cal{N}}/2\le S^z\le{\cal{N}}$, 
see Ref.~\onlinecite{linear_ind}, 
i.e., all these states contribute to the partition function of the spin system.
Due to their huge degeneracy 
they dominate the low-temperature thermodynamics for magnetic fields $h$ near $h_1$,
see Refs.~\onlinecite{d&r_b,d&r_c,d&r_d} and Sec.~\ref{sec5}.

Now we extend the above procedure considering the interacting localized-magnon states
to calculate the degeneracy of the low-lying excited states.
In the language of hard-core objects 
the hard-core rule is partially relaxed and the hard-core objects may partially overlap, 
see for illustration Fig.~\ref{fig3}(b).
In what follows we call them overlapping hard-core states.
Note that a complete overlap 
(i.e., two hard-core objects on the same site) 
is strictly forbidden, 
since a corresponding spin state does not exist.
First we consider the lowest excited states in the sectors ${\cal{N}}/2\le S^z<{\cal{N}}-1$
(i.e., ${\cal{N}}/2 \ge n > 1$). 
These states have two and only two neighboring singlets, 
i.e.,
(i) we have only one  pair ($\nu=1$) of neighboring vertical singlets and 
(ii) in the  hard-core model two (and only two) hard-core objects overlap, 
cf. Fig.~\ref {fig3}(b).
We denote the degeneracy of the first excited states in the $n$-magnon subspace by $x_{{\cal{N}}}(n)$.
Clearly,
$x_{{\cal{N}}}(n)$ for $n=2,\ldots,{\cal{N}}/2$ 
equals the canonical partition function 
of a system with $n-2$ non-overlapping hard-core objects 
and one composite hard-core object built by two overlapping objects.
In the one-dimensional case we immediately conclude,
that $x_{{\cal{N}}}(n)={\cal{N}}Z_{{\rm{hc}}}(n-2,{\cal{N}}-4)$,
where $Z_{{\rm{hc}}}(m,{\cal{M}})$ is the canonical partition function 
of $m$ hard dimers on ${\cal{M}}$-site chain, 
however, (instead of periodic) with open boundary conditions.
In the two-dimensional case 
$x_{{\cal{N}}}(n)=2{\cal{N}}\tilde{Z}_{{\rm{hc}}}(n-2,{\cal{N}}-8)$,
where $\tilde{Z}_{{\rm{hc}}}(m,{\cal{M}})$ is the canonical partition function 
of $m$ hard squares on a (periodic) ${\cal{M}}$-site square lattice 
with a ``dumbbell void'' oriented either in horizontal or vertical direction, 
where the two overlapping hard squares are located.
Using similar reasonings we may find the degeneracy of higher excited states with $\nu>1$.
However,
since we are interested in thermodynamic properties,
we need in fact the energies and degeneracies of excited states only in certain combinations 
which enter the lattice-gas thermodynamics, see Sec.~\ref{sec6}.

It is straightforward to determine the ground-state degeneracy 
$g_{\cal{N}}(n)$ in the remaining subspaces, i.e., for $S^z={\cal{N}}/2-1,\ldots,0$
(that is for $n={\cal{N}}/2+1,\ldots,{\cal{N}}$)
and
the first-excited-state degeneracy 
$x_{\cal{N}}(n)$ in the subspaces with $S^z={\cal{N}}/2-1,\ldots,2$
(that is for $n={\cal{N}}/2+1,\ldots,{\cal{N}}-2$). 
The ground state is then a state with minimal overlap of hard-core objects, 
e.g., one sublattice is completely occupied 
and the other one carries $n-{\cal{N}}/2$ hard-core objects. 
Then the first excited state is a state with one (and only one) empty site in one sublattice 
together with a neighboring empty site on the other sublattice (``composite hole'').    
As a result, 
one has a simple ``particle-hole'' symmetry for hard-core objects:
$g_{\cal{N}}(n)=g_{\cal{N}}({\cal{N}}-n)$
and
$x_{\cal{N}}(n)=x_{\cal{N}}({\cal{N}}-n)$.
Note that such a particle-hole symmetry is valid also for higher excited states.

We can find the degeneracies $g_{\cal{N}}(n)$ and $x_{\cal{N}}(n)$ analytically in the one-dimensional case
knowing the partition functions ${\cal{Z}}_{{\rm{hc}}}(n,{\cal{N}})$ and $Z_{{\rm{hc}}}(m,{\cal{M}})$.
These quantities follow from the grand-canonical partition function
for the one-dimensional hard-dimer model with periodic and open boundary conditions
through calculation of derivatives with respect to the hard-dimer activity $z$ at $z=0$,
e.g.,
$n!{\cal{Z}}_{{\rm{hc}}}(n,{\cal{N}})=d^n\Xi_{{\rm{pbc}}}(z,{\cal{N}})/dz^n\vert_{z=0}$.
The grand-canonical partition function for the one-dimensional hard-dimer model
can be obtained by the transfer-matrix method, see, e.g., Ref.~\onlinecite{baxter}.
For periodic boundary conditions imposed we have
\begin{eqnarray}
\Xi_{{\rm{pbc}}}(z,{\cal{N}})=\lambda_+^{{\cal{N}}}+\lambda_-^{{\cal{N}}}
\label{4.01}
\end{eqnarray}
with $\lambda_{\pm}=(1\pm\sqrt{1+4z})/2$.
For open boundary conditions imposed we have
\begin{eqnarray}
\Xi_{{\rm{obc}}}(z,{\cal{N}})
=\left(a_+^2 + 2\sqrt{z}a_+b_+ + zb_+^2\right)\lambda_+^{{\cal{N}}-1}
\nonumber\\
+\left(a_-^2 + 2\sqrt{z}a_-b_- + zb_-^2\right)\lambda_-^{{\cal{N}}-1}
\label{4.02}
\end{eqnarray}
with the same $\lambda_{\pm}$
and
$a_{\pm}=(1\pm\sqrt{1+4z})/\sqrt{2C_{\pm}}$,
$b_{\pm}=\sqrt{2z/C_{\pm}}$,
$C_{\pm}=1+4z\pm\sqrt{1+4z}$.
In the two-dimensional case, the required partition functions can be easily found numerically.
For example, 
for ${\cal N} = N/2=16$ we get
${\cal{Z}}_{\rm{hc}}(n,{\cal N})=1,16,88,208,228,128,56,16,2$
for $n=0,1,2,3,4,5,6,7,8$
or
$2 {\cal N}\tilde{Z}_{{\rm{hc}}}(n-2,{\cal N}-8)=32,256,576,448,64$ 
for $n=2,3,4,5,6$.

To check the hard-core predictions 
we have calculated the degeneracies of the ground states and the lowest excitations 
as well as the excitation gaps 
of the $s=1/2$ frustrated two-leg ladder and bilayer
by full diagonalization for finite spin systems up to $N=32$
and various sectors of total $S^z$.
The exact diagonalization data coincide perfectly 
with the corresponding data obtained by analytical formulas
(\ref{3.02}), (\ref{3.03}), (\ref{3.04}), (\ref{4.01}), and (\ref{4.02})
and numerics for ${\cal{Z}}_{\rm{hc}}(n,{\cal{N}})$ and $\tilde{Z}_{{\rm{hc}}}(m,{\cal{M}})$.

As mentioned in Sec.~\ref{sec3},
at the fields $h_1=\epsilon_1$
all independent and at  $h_2=J_2$ all interacting localized-magnon ground states are degenerate
which leads to a jump in the magnetization curve,
see Fig.~\ref{fig2}.
These degeneracies of localized-magnon states at $h_1$ and $h_2$,
${\cal W}_1=\sum_{n=0}^{{ \cal N}/2}g_{\cal{N}}(n)$ 
and
${\cal W}_2=\sum_{n={\cal N }/2}^{{\cal N }}g_{\cal N }(n)$,
grow exponentially with the system size $N$.
Due to the particle-hole
symmetry we have ${\cal W }_1={\cal W }_2 ={\cal W}$.
The exponential growth of ${\cal{W}}$ 
leads to a nonzero ground-state residual entropy 
$S(T=0,h,N)/N=(\ln{\cal{W}})/N \ne 0$ at $h=h_1$ and $h=h_2$.

To summarize Secs.~\ref{sec3} and \ref{sec4},
we have characterized the low-energy eigenstates of the frustrated two-leg ladder and bilayer 
in the strong-coupling regime
calculating their energies and degeneracies.
In the next sections we show 
how due to a simple structure of these low-energy eigenstates
their contributions to thermodynamics can be obtained with the help of auxiliary lattice-gas models.

\section{Lattice-gas models with nearest-neighbor exclusion (hard-core
models)}
\label{sec5}

We want to calculate a partition function $Z(T,h,N)$ of the spin system (\ref{2.01}).
In a first step we consider in this  section 
the contribution of independent localized-magnon states 
(non-overlapping hard-core states) 
to the partition function. 
Recall that these states are the ground states in the subspaces with $S^z=N/2,\ldots,N/4$
with energies $E^{{\rm{lm}}}_n(h)$, $n=N/2-S^z=0,1,\ldots,N/4$
and degeneracies $g_{\cal{N}}(n)$.
If the magnetic field $h$ is around the saturation filed $h_1$ 
they will give the dominant contribution at low temperatures $T$.
Therefore
\begin{eqnarray}
\label{5.01}
Z(T,h,N)
&\approx&
Z_{{\rm{lm}}}(T,h,N)
=\sum_{n=0}^{\frac{N}{4}}g_{\cal{N}}(n)e^{-\frac{E_n^{\rm{lm}}(h)}{T}}
\nonumber\\
&=&e^{-\frac{E_{{\rm{FM}}}-h\frac{N}{2}}{T}}
\sum_{n=0}^{\frac{N}{4}}g_{\cal{N}}(n)e^{\frac{\mu}{T}n},
\end{eqnarray}
where $\mu=\epsilon_1-h=h_1-h$. 
Obviously, 
the magnetic field and the temperature enter the thermodynamic quantities within the hard-core description
via the combination $(h_1-h)/T$, only.
Since $g_{\cal{N}}(n)$ is the canonical partition function ${\cal{Z}}_{{\rm{hc}}}(n,{\cal{N}})$
of $n$ hard dimers on a chain of ${\cal{N}}$ sites
or
of $n$ hard squares on a square lattice of ${\cal{N}}$ sites,
$\Xi_{{\rm{hc}}}(T,\mu,{\cal{N}})
=\sum_{n=0}^{{\cal{N}}/2}g_{\cal{N}}(n)e^{\mu n/T}$
is the grand-canonical partition function
of the corresponding one-dimensional hard-dimer model or two-dimensional hard-square model
and $\mu$ is the chemical potential of the hard-core objects.
It is also useful to rewrite $Z_{{\rm{lm}}}(T,h,N)$ in the following form:
\begin{eqnarray}
\label{5.02}
Z_{{\rm{lm}}}(T,h,N)
=e^{-\frac{E_{{\rm{FM}}}-h\frac{N}{2}}{T}}
\nonumber\\
\times
\sum_{n_1=0,1}
\ldots
\sum_{n_{\cal{N}}=0,1}
e^{\frac{\mu}{T}\sum_{m=1}^{{\cal{N}}}n_m}R(\{n_m\}),
\end{eqnarray}
where the factor $R(\{n_m\})$ takes care about the hard-core rule,
i.e., 
it is $0$ if the spatial configuration $\{n_m\}$ violates the hard-core rule
but it is $1$ if the hard-core rule is fulfilled.
For example, for the one-dimensional hard dimers
$R(\{n_m\})=(1-n_1n_2)(1-n_2n_3)\ldots(1-n_{{\cal{N}}-1} n_{{\cal{N}}})(1-n_{{\cal{N}}}n_1)$.

In summary,
we arrive at the basic relation for the independent localized-magnon state contribution
to the Helmholtz free energy of the spin system (\ref{2.01})
\begin{eqnarray}
\label{5.03}
\frac{F_{{\rm{lm}}}(T,h,N)}{N}
=\frac{E_{{\rm{FM}}}}{N}-\frac{h}{2}-\frac{T}{2}\frac{\ln\Xi_{{\rm{hc}}}(T,\mu,{\cal{N}})}{{\cal{N}}}.
\end{eqnarray}
The entropy $S$, the specific heat $C$, the (uniform) magnetization $M$,
and the (uniform) susceptibility $\chi$
follows from (\ref{5.03})
according to usual formulas,
$S(T,h,N)=-\partial F(T,h,N)/\partial T$,
$C(T,h,N)=T\partial S(T,h,N)/\partial T$,
$M(T,h,N)=N/2-\overline{n}$,
$\overline{n}=T\partial \ln\Xi(T,\mu,{\cal{N}})/\partial\mu$,
$\chi(T,h,N)=\partial M(T,h,N)/\partial h=\partial\overline{n}/\partial\mu$.

To examine the ordering of hard-core objects (localized magnons)
we consider the average total numbers of hard-core objects on the sublattices $A$ and $B$,
$\overline{n_A}$ and $\overline{n_B}$.
Obviously $\overline{n}=\overline{n_A}+\overline{n_B}$,
whereas the value of the difference $\vert \overline{n_A}-\overline{n_B} \vert$ may play a role of the order parameter $m$.
Introducing an infinitesimally small symmetry-breaking staggered component into the chemical potential,
i.e.,
$\mu\to\mu_A=\mu+\delta\mu$ on the sublattice $A$
and
$\mu\to\mu_B=\mu-\delta\mu$ on the sublattice $B$,
$\mu=h_1-h$, $\delta\mu=-\delta h$,
we calculate the staggered magnetization
$M_{\rm{st}}(T,h,\delta h,N)=M_A-M_B=-\overline{n_A}+\overline{n_B}
=-T\partial \ln\Xi(T,\mu_A,\mu_B,{\cal{N}})/\partial\mu_A
+T\partial \ln\Xi(T,\mu_A,\mu_B,{\cal{N}})/\partial\mu_B
=\chi_{\rm{st}}(T,h,N) \delta h$,
where $\chi_{\rm{st}}=\partial M_{\rm{st}}/\partial\delta h$
is the staggered susceptibility.
Decreasing the temperature 
a divergence of the staggered susceptibility in the thermodynamic limit
signals a transition to an ordered phase, 
where the symmetry of the occupation of both sublattices with localized magnons (hard-core objects)
can be spontaneously broken.

Below we discuss briefly thermodynamic quantities 
as they follow from the lattice-gas models with nearest-neighbor exclusion
and compare them with exact diagonalization data 
for the frustrated quantum Heisenberg antiferromagnets on finite lattices.

\subsection{Frustrated two-leg ladder}
\label{sec5a}

We begin with the case of the frustrated two-leg ladder.
Using the transfer-matrix result for one-dimensional hard dimers [see Eq. (\ref{4.01})],
\begin{eqnarray}
\Xi_{{\rm{hc}}}(T,\mu,{\cal{N}})
=\lambda_+^{\cal{N}} + \lambda_-^{\cal{N}},
\nonumber\\
\lambda_{\pm}=\frac{1}{2}\pm\sqrt{\frac{1}{4}+ z} \; , \; z=e^{\frac{\mu}{T}}
\; , \;\mu=h_1-h\; , \;
\label{5.04}
\end{eqnarray}
one can easily find all thermodynamic quantities,
$S$, $C$, $M$, and $\chi$,
see Eqs. (\ref{A.01}) -- (\ref{A.04}) and (\ref{A.06}) -- (\ref{A.09}) in
the Appendix \ref{a}.
The main features of the low-temperature thermodynamic behavior of the frustrated two-leg ladder in this regime
are as follows:
(i) the jump in zero-temperature magnetization  at $h=h_1$, 
cf. Fig.~\ref{fig2}, 
is smeared out at low but finite nonzero temperatures;
(ii) the entropy $S(T,h_1,N)/N$ remains finite 
and approaches $S(T=0,h_1,N)/N=(1/2)\ln\varphi$ as $T\to 0$, 
where $\varphi=(1+\sqrt{5})/2$ is the golden mean
[note that due to the particle-hole symmetry explained in Sec.~\ref{sec4} 
there is the same  ground-state residual entropy at $h=h_2$,
$S(T=0,h_2,N)/N=(1/2)\ln\varphi$];
(iii) the specific heat shows an extra low-temperature maximum if $h$ slightly deviates from $h_1$ 
indicating a new low-energy scale settled by the set of independent localized-magnon states.
A comprehensive analysis of low-temperature high-field thermodynamic quantities $S$, $C$, $M$, and $\chi$
based on the hard-dimer description (\ref{5.03}), (\ref{5.04}) can be found in Refs.~\onlinecite{d&r_b,d&r_d}.

To calculated the staggered susceptibility 
(not considered in previous papers\cite{d&r_b,d&r_d}) 
we have to consider different chemical potentials on the sublattices.
Then the grand-canonical partition function reads
\begin{eqnarray}
\label{5.05}
\Xi_{{\rm{hc}}}(T,\mu_A,\mu_B,{\cal{N}})
=\xi_+^{\frac{{\cal{N}}}{2}} + \xi_-^{\frac{{\cal{N}}}{2}}, \quad \quad \quad  
\\
\xi_{\pm}=\frac{1}{2}+\frac{z_A+z_B}{2}
\pm\sqrt{\frac{1}{4}+\frac{z_A+z_B}{2}+\frac{\left(z_A-z_B\right)^2}{4}},
\nonumber\\
z_A=e^{\frac{\mu_A}{T}}, 
\;
z_B=e^{\frac{\mu_B}{T}},
\; 
\mu_A=h_1-h_A,
\; 
\mu_B=h_1-h_B.
\nonumber
\end{eqnarray}
Eq. (\ref{5.05}) immediately yields the staggered susceptibility $\chi_{{\rm{st}}}(T,h,N)$,
see Eqs. (\ref{A.05}), (\ref{A.10}).
As expected, 
for the one-dimensional problem there is no divergence at $T>0$.
However, 
for $h \le h_1$ the staggered susceptibility diverges at $T=0$.
Precisely at $h = h_1$ one finds
$\chi_{{\rm{st}}}(T,h_1,N)/{\cal{N}} \to (1/\sqrt{5}) \; T^{-1} \approx 0.447\,214\;T^{-1} $. 
For $h < h_1$ one finds 
$\chi_{{\rm{st}}}(T,h,N)/{\cal{N}} = ({\cal{N}}/4)\; T^{-1} $ 
for finite $\cal{N}$.
In the thermodynamic limit $\cal{N} \to \infty $ the divergence becomes exponential 
$\chi_{{\rm{st}}}(T,h,N)/{\cal{N}} = (1/2) T^{-1}  e^{(h_1-h)/(2T)}$.
Note 
that this temperature dependence is identical to that of the Ising chain.
Note further that, trivially, $T\chi_{{\rm{st}}}(T,h,N)/{\cal{N}}$ goes to zero at $T=0$ 
if $h>h_1$.

\subsection{Frustrated bilayer}
\label{sec5b}

We turn to the case of the frustrated bilayer.
By contrast to the one-dimensional case discussed in the previous section
there is no exact analytical solution 
for the corresponding two-dimensional hard-square model.
Thermodynamic functions for the hard-square model can be obtained by direct computations
only if ${\cal{N}}$ is small enough (see Appendix \ref{b}).
For larger  ${\cal{N}}$ we use classical Monte Carlo simulations\cite{footnote2} 
(see also Appendix \ref{b}).

First we illustrate the validity of the hard-square description
by comparison with exact diagonalization data for finite systems.
For that we show in Fig.~\ref{fig4} the specific heat $C(T,h,N)$.
\begin{figure}
\begin{center}
\includegraphics[clip=on,width=7.5cm,angle=0]{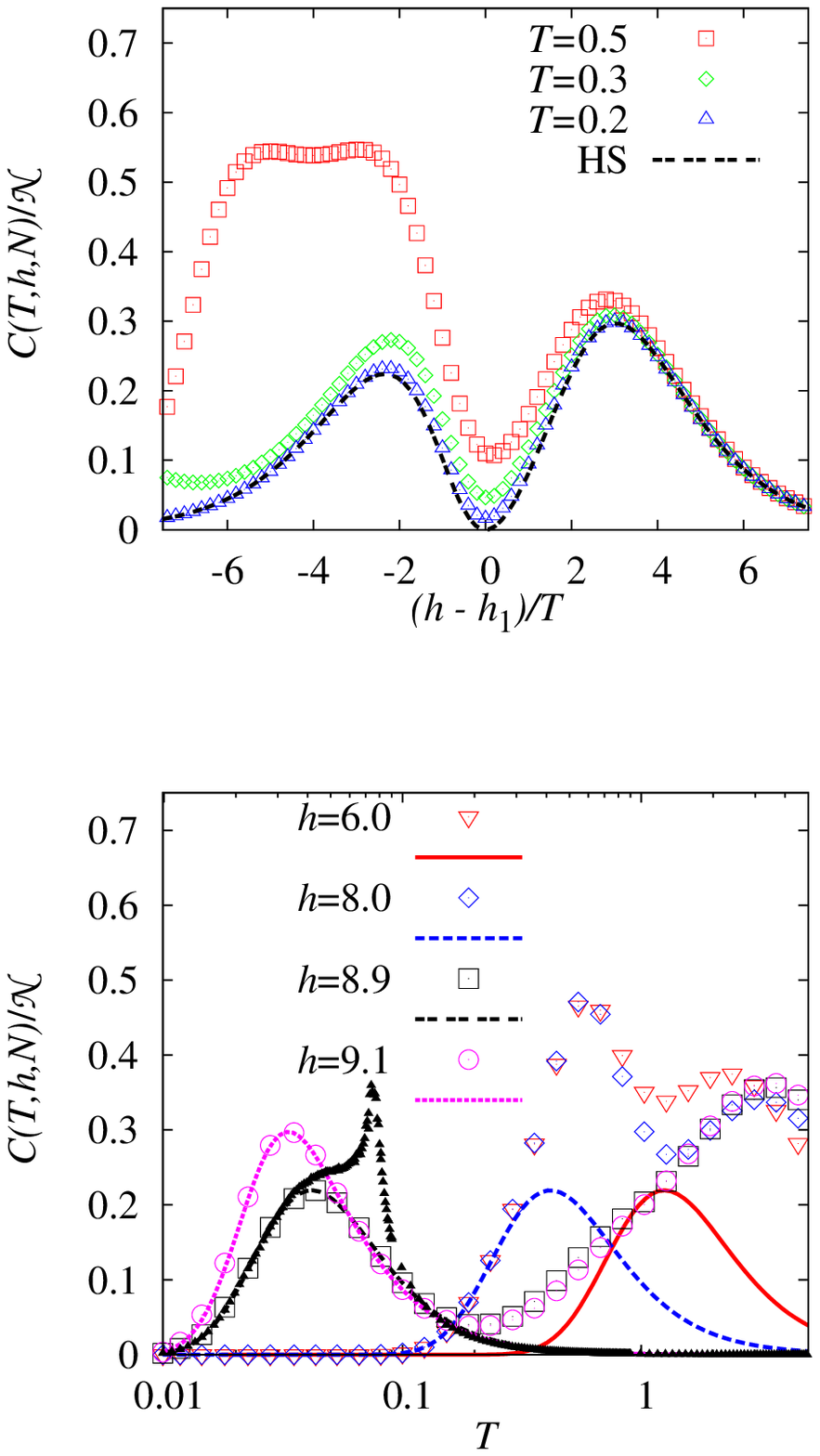}
\caption
{(Color online)
Specific heat $C$ for the frustrated bilayer with $J_1=1$, $J_2=5$ ($h_1=9$
and $h_2=5$):
Exact diagonalization data versus hard-square predictions.
Upper panel: 
$C$ in dependence on the hard-core parameter $(h-h_1)/T$, cf. Eq.~(\ref{5.01})
[symbols:
spin system with $N=20$ (${\cal N}=10$), 
double-dashed line: 
hard-square model with ${\cal N}=10$].  
Lower panel: 
$C$ in dependence on the temperature $T$ for various values of magnetic field $h$ 
[open symbols:
spin system with $N=16$ (${\cal N}=8$), 
lines: 
hard-square model with ${\cal N}=8$, 
black filled up-triangles: 
Monte-Carlo results for the hard-square model for large ${\cal N}$ up to $800 \times 800$ for $h=8.9$].}
\label{fig4}
\end{center}
\end{figure}
According  to the upper panel of  Fig.~\ref{fig4} 
for $T \lesssim 0.2$ the specific heat depends only on the hard-core parameter $(h-h_1)/T$ 
and the hard-core description is valid over the entire range of magnetic fields.
The temperature dependence of $C$ for various magnetic fields is shown in the lower panel of Fig.~\ref{fig4}. 
It is again obvious that the low-temperature behavior is well described by the hard-square model.
However, 
the temperature range of the validity of the hard-square model becomes smaller with increasing $h_1-h$. 
In particular, 
the position and the height of the characteristic extra low-temperature maximum in $C$ 
is correctly described only if $h_1-h \lesssim 0.5$.

Similar as for the one-dimensional case at $h=h_1$,
there is a ground-state residual entropy $S(T=0,h_1,N)/N = 0.2037...=(1/2)\ln\kappa(1)$.
This value follows from the hard-square model theory,
which predicts for the hard-square entropy constant
$\kappa(1)=1.50304808\ldots$.\cite{baxter_99}
Moreover, due to the particle-hole symmetry, see 
Sec.~\ref{sec4},
there is the same ground-state residual entropy  at $h=h_2$,
$S(T=0,h_2,N)/N = (1/2)\ln\kappa(1)$.

The main peculiarity of the low-temperature thermodynamics of the frustrated bilayer around $h_1$ 
is connected with an order-disorder phase transition 
which is inherent in the hard-square model at $T_c\approx (h_1-h)/1.3340$.\cite{baxter_hs,d&r_c}
The phase transition has pure geometrical origin:
if the density of hard squares [controlled by the activity
$z=e^{(h_1-h)/T}$] increases 
they start to occupy only one of two sublattices.
The  critical value of the activity is $z_c=3.7962\ldots$.
The universality class is that of the two-dimensional Ising model,
i.e.,  we have a logarithmic singularity for the specific heat $C\propto\ln\vert T-T_c\vert$
and critical indices 
$\beta=1/8$ for the order parameter 
[$m \propto (T_c-T)^{\beta}$, $T<T_c$]
and 
$\gamma=7/4$ for the staggered susceptibility 
[$\chi_{\rm{st}}\propto \vert T-T_c \vert^{-\gamma}$].
This conclusion drawn from the classical hard-square model 
taking into account only independent localized-magnon states (non-overlapping hard-square states)
is reliable only for a quite small interval $(h_1-h)/h_1 \ll 1$.  
However, 
in the next section we will demonstrate 
that the range of validity of the classical description can be significantly extended 
including interacting localized-magnon states (overlapping hard-square states).

\section{Lattice-gas models with finite repulsion}
\label{sec6}

Taking into account also the interacting localized-magnon states, 
i.e., low-lying excitations,
we will present a significantly  improved lattice-gas description 
of the low-temperature thermodynamics of the considered spin models in this section. 
For that we use the information on the energies and degeneracies of these excitations, 
given in Secs.~\ref{sec3} and \ref{sec4}.
We start with the partition function for hard-core objects $Z_{{\rm{lm}}}(T,h,N)$ 
and use its form given in Eq.~ (\ref{5.02}).
The hard-core rule is taken into account by the factor $R(\{n_m\})$.
To relax this rule we can preserve the form of the partition function, 
but we have to replace $R(\{n_m\})$ 
by a modified factor $e^{-(V/T)\sum_{(mp)}n_mn_p}$
taking into account the excitation energy $V=J_1$.    
Then we arrive at the following formula instead of Eq.~(\ref{5.02})
\begin{eqnarray}
\label{6.01}
Z(T,h,N)
\approx
Z_{{\rm{LM}}}(T,h,N)
=e^{-\frac{E_{{\rm{FM}}}-h\frac{N}{2}}{T}}
\nonumber\\
\times
\sum_{n_1=0,1}
\ldots
\sum_{n_{\cal{N}}=0,1}
e^{\frac{\mu}{T}\sum_{m=1}^{{\cal{N}}}n_m}
e^{-\frac{V}{T}\sum_{(mp)}n_mn_p},
\end{eqnarray}
where $\mu=h_1-h$ and the sum $\sum_{(mp)}$ runs over all nearest-neighbor bonds on the underlying lattice.
The limit $V/T\to\infty$ now corresponds to the hard-core limit given in Eq. (\ref{5.02}), 
since for $V/T\to\infty$ we get $e^{-(V/T)\sum_{(mp)}n_mn_p}\to R(\{n_m\})$, 
i.e., the excitations get zero statistical weight.  
Note that in the improved lattice-gas description 
there is now an explicit temperature dependence in addition to the hard-core combination $(h_1-h)/T$.
Apart from the trivial  factor $e^{-(E_{{\rm{FM}}}-hN/2)/T}$
the partition function $Z_{{\rm{LM}}}(T,h,N)$ in (\ref{6.01})
is the grand-canonical partition function $\Xi_{{\rm{lg}}}(T,\mu,{\cal{N}})$
of the lattice-gas model with finite nearest-neighbor repulsion $0<V<\infty$.
Instead of Eq. (\ref{5.03}) we now have
\begin{eqnarray}
\label{6.02}
\frac{F_{{\rm{LM}}}(T,h,N)}{N}
=\frac{E_{{\rm{FM}}}}{N}-\frac{h}{2}-\frac{T}{2}\frac{\ln\Xi_{{\rm{lg}}}(T,\mu,{\cal{N}})}{{\cal{N}}}
\end{eqnarray}
with
\begin{eqnarray}
\label{6.03}
\Xi_{{\rm{lg}}}(T,\mu,{\cal{N}})
=\sum_{n_1=0,1}\ldots\sum_{n_{{\cal{N}}}=0,1}
e^{-\frac{{\cal{H}}(\{n_m\})}{T}} ,
\end{eqnarray}
where
\begin{eqnarray}
\label{6.04}
{\cal{H}}(\{n_m\})
=\sum_{m=1}^{{\cal{N}}}
\left(-\mu n_{m}+Vn_{m}n_{m+1}\right)
\end{eqnarray}
in the one-dimensional case 
or
\begin{eqnarray}
\label{6.05}
{\cal{H}}(\{n_m\})
=\sum_{m_x=1}^{{\cal{N}}_x}\sum_{m_y=1}^{{\cal{N}}_y}
\left[-\mu n_{m_xm_y}
\right.
\nonumber\\
\left.
+V\left(n_{m_xm_y}n_{m_x+1,m_y}+n_{m_xm_y}n_{m_x,m_y+1}\right)
\right]
\end{eqnarray}
in the two-dimensional case.
From $\Xi_{{\rm{lg}}}(T,\mu,{\cal{N}})$ 
the  thermodynamic quantities can be found in usual way, 
cf. Appendices \ref{a} and \ref{b}.

The following remarks are pertinent.
Firstly,
we notice that the initial quantum spin model (\ref{2.01}) has $2^N$ states
and obviously not all of them are included in the effective models.
Thus, the hard-core models contain
either $\varphi^{N/2}\approx 1.272^N$ states (one-dimensional case)
or $\kappa(1)^{N/2}\approx 1.226^N$ states (two-dimensional case).
The lattice-gas model with finite repulsion has $2^{N/2}\approx 1.414^N$ states.

Secondly,
we note that the particle-hole symmetry has some useful consequences.
After making the transformation
$n_m\to\tilde{n}_m=1-n_m$ in Eq. (\ref{6.04})
or
$n_{m_xm_y}\to\tilde{n}_{m_xm_y}=1-n_{m_xm_y}$ in Eq. (\ref{6.05})
we arrive at the Hamiltonian
${\cal{H}}(\{\tilde{n}_m\})$
with $-\mu+2V$ instead of $\mu$ and shifted by ${\cal{N}}(-\mu+V)$ in the case (\ref{6.04})
or
with $-\mu+4V$ instead of $\mu$ and shifted by ${\cal{N}}(-\mu+2V)$ in the case (\ref{6.05}).
This fact implies,
that
$\Xi_{\rm{lg}}(T,\mu,{\cal{N}})=e^{{\cal{N}}(\mu-V)/T}\Xi_{\rm{lg}}(T,-\mu+2V,{\cal{N}})$
in the one-dimensional case
or
$\Xi_{\rm{lg}}(T,\mu,{\cal{N}})=e^{{\cal{N}}(\mu-2V)/T}\Xi_{\rm{lg}}(T,-\mu+4V,{\cal{N}})$
in the two-dimensional case.
In particular,
this yields identical ground-state residual entropies at the fields $h_1$ and $h_2$.
Moreover,
the lattice-gas model with finite repulsion provides similar descriptions of the initial quantum spin model
around both characteristic fields $h_1$ and $h_2$.

Thirdly,
it is useful to introduce the on-site spin variables $\sigma=\pm 1$
related to the site occupation numbers $n=0,1$ as follows:
$\sigma=2n-1$ and $n=(1+\sigma)/2$.
Then Eq. (\ref{6.04}) becomes the Hamiltonian
of the antiferromagnetic Ising chain in a uniform magnetic field
\begin{eqnarray}
{\cal{H}}&=&{\cal{N}}\left(-\frac{\mu}{2}+\frac{V}{4}\right)
+\sum_{m=1}^{{\cal{N}}}
\left(-{\Gamma} \sigma_{m}+{\cal{J}}\sigma_{m}\sigma_{m+1}\right),
\nonumber\\
{\Gamma}&=&\frac{\mu}{2}-\frac{V}{2},
\; 
\mu= h_1-h , 
\;
h_1=2J_1+J_2,
\;
{\cal{J}}=\frac{V}{4}>0,
\label{6.06}
\end{eqnarray}
whereas Eq. (\ref{6.05}) becomes the Hamiltonian
of the square-lattice antiferromagnetic Ising model in a uniform magnetic field
\begin{eqnarray}
{\cal{H}}&=&{\cal{N}}\left(-\frac{\mu}{2}+\frac{V}{2}\right)
+\sum_{m_x=1}^{{\cal{N}}_x}\sum_{m_y=1}^{{\cal{N}}_y}
\left[
-{\Gamma} \sigma_{m_xm_y}
\right.
\nonumber\\
&& \left.
+{\cal{J}}\left(\sigma_{m_xm_y}\sigma_{m_x+1,m_y}+\sigma_{m_xm_y}\sigma_{m_x,m_y+1}\right)
\right],
\nonumber\\
{\Gamma}&=&\frac{\mu}{2}-V,
\; 
\mu= h_1-h , 
\;
h_1=4J_1+J_2,
\;
{\cal{J}}=\frac{V}{4}>0.
\label{6.07}
\end{eqnarray}
Let us mention again that for $V=J_1$ we get correspondence to initial quantum spin systems. 
Note further, 
that the residual entropy present in the initial quantum spin systems at $h=h_1$ and $h=h_2$ 
corresponds to the known residual entropy of the Ising antiferromagnet 
at 
$\Gamma = \pm 2 {\cal J}$ (one-dimensional case) 
and
$\Gamma = \pm 4 {\cal J}$ (two-dimensional case).\cite{metcalf}
From Ref.~\onlinecite{metcalf} we know 
that the ground state entropy per site at the critical fields is
$\ln[(1+\sqrt{5})/2]=0.4812\ldots$ (one-dimensional case)
and
$\approx 0.4075$ (two-dimensional case)
that coincides with the corresponding data for the considered quantum spin systems
reported in Secs.~\ref{sec5a} and \ref{sec5b}.

Now we discuss the low-temperature properties of the quantum spin models
under consideration 
on the basis of Eqs. (\ref{6.02}) -- (\ref{6.07})
considering separately the frustrated two-leg ladder and the frustrated bilayer.

\subsection{Frustrated two-leg ladder}
\label{sec6a}

The one-dimensional lattice-gas model with finite nearest-neighbor repulsion admits rigorous analysis.
With the help of the transfer-matrix method we get
\begin{eqnarray}
\label{6.08}
\Xi_{{\rm{lg}}}(T,\mu,{\cal{N}})=\lambda_+^{{\cal{N}}}+\lambda_-^{{\cal{N}}},
\nonumber\\
\lambda_{\pm}
=\frac{1}{2}+\frac{1}{2}ze^{-\frac{V}{T}}
\pm
\sqrt{\left(\frac{1-ze^{-\frac{V}{T}}}{2}\right)^2+z} 
\end{eqnarray}
with $z=e^{(h_1-h)/T}$.
Thermodynamic quantities for finite and infinite systems are given in Appendix \ref{a},
Eqs. (\ref{A.11}) -- (\ref{A.14}) and Eqs. (\ref{A.16}) -- (\ref{A.19}).
Obviously $\lambda_{\pm}$ in Eq. (\ref{6.08}) transforms into $\lambda_{\pm}$ in Eq. (\ref{5.04})
if $V/T\to\infty$.
It is interesting to note that according to Eq. (\ref{6.08})
$\lambda_{\pm}(\mu=2V)=e^{V/T}\lambda_{\pm}(\mu=0)$
and hence
$\Xi_{{\rm{lg}}}(T,\mu=2V,{\cal{N}})=e^{{\cal{N}}V/T}\Xi_{{\rm{lg}}}(T,\mu=0,{\cal{N}})$.
This relation was mentioned already on the basis of particle-hole symmetry for hard-core objects.

The transfer-matrix calculation in the case of different chemical potentials
$\mu_A$ and $\mu_B$  on the sublattices $A$ and $B$
leads to the following result for the grand-canonical partition function
\begin{eqnarray}
\label{6.09}
\Xi_{{\rm{lg}}}(T,\mu_A,\mu_B,{\cal{N}})
=\xi_+^{\frac{{\cal{N}}}{2}} + \xi_-^{\frac{{\cal{N}}}{2}},
\nonumber\\
\xi_{\pm}^2-\left(1+z_A+z_B+z_Az_Be^{-\frac{2V}{T}}\right)\xi_{\pm}
\nonumber\\
+z_Az_B\left(1-e^{-\frac{V}{T}}\right)^2=0.
\end{eqnarray}
$\xi_{\pm}$ in Eq. (\ref{6.09}) transforms into $\xi_{\pm}$ in Eq. (\ref{5.05})
if $V/T\to\infty$.
With  (\ref{6.09}) we can calculate the staggered susceptibility $\chi_{\rm{st}}(T,h,N)$,
see Eqs. (\ref{A.15}) and (\ref{A.20}) in Appendix \ref{a}.

We start with a general discussion of the low-temperature properties of the frustrated two-leg ladder
based on its correspondence to the Ising chain (\ref{6.06}). 
The one-dimensional Ising antiferromagnet 
exhibits antiferromagnetic long-range order along the line $T=0$ if $\vert {\Gamma}\vert <2{\cal{J}}$
and ferromagnetic long-range order along the line $T=0$ if $\vert {\Gamma}\vert >2{\cal{J}}$,
whereas for any nonzero temperature it is in a disordered phase.
In lattice-gas language this means that at $T=0$
the lattice is empty
when $\mu<0$,
one sublattice is completely occupied and the other one is empty (two-fold degenerate phase)
when $0<\mu<2V$,
and all lattice sites are occupied 
when $2V<\mu$.
In terms of the initial quantum Heisenberg ladder this means 
that at $T=0$ the Ising-like antiferromagnetic long-range ordered phase 
occurs if $h_2=h_1-2J_1<h<h_1$ only,
i.e., for magnetic fields within the one-half magnetization plateau.
Thermal fluctuations destroy perfect orders
and a smooth crossover
from the empty lattice to the lattice occupied by ${\cal{N}}$ localized magnons
takes place at any fixed nonzero temperature
as $h$ decreases from above the saturation field $h_1$ to zero.

Now we turn to numerics for finite systems.
We fix $J_1=1$ and set for concreteness $J_2=4>J_2^{c}\approx 3.00J_1$.
Note that with increasing of $J_2$ 
the lattice-gas description is expected to become better, 
since excitations 
not described by the lattice-gas model 
are shifted to higher energies.
In Fig.~\ref{fig5}
we compare some results for the specific heat of finite systems
obtained from exact diagonalization of the spin systems 
and from the lattice-gas formulas,
see Eqs. (\ref{6.02}), (\ref{6.08}), (\ref{6.09}) and (\ref{A.11}) -- (\ref{A.20})
with $V=J_1=1$. 
For illustration we show in the upper panel also the hard-dimer result
($V \to \infty$).
\begin{figure}
\begin{center}
\includegraphics[clip=on,width=7.5cm,angle=0]{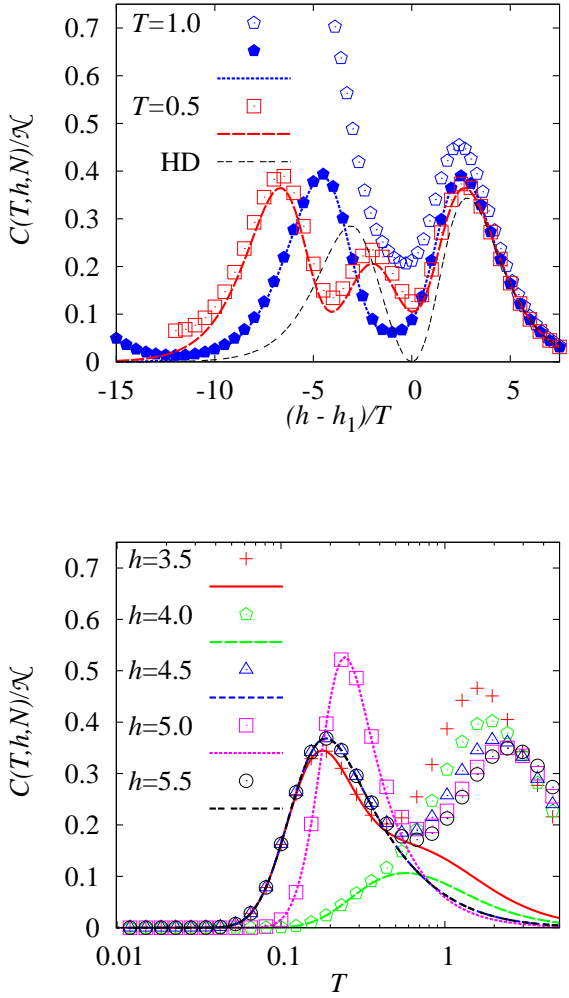}
\caption
{(Color online)
Specific heat $C$ for the frustrated two-leg ladder  
with $J_1=1$, $J_2=4$ ($h_1=6$, $h_2=4$) and $J_1=1$, $J_2=10$ ($h_1=12$, $h_2=10$):
Exact diagonalization data for $N=16$ (${\cal N}=8$) 
versus 
lattice-gas predictions for ${\cal{N}}=8$.
Upper panel: 
$C$ in dependence on the hard-core parameter $(h-h_1)/T$ for $T=0.5$ and $T=1$
[open symbols:
spin system with $J_2=4$, 
filled pentagons: 
spin system with $J_2=10$,
long-dashed and dotted lines:
lattice-gas model, 
double-dashed line:
hard-dimer model].
Lower panel: 
$C$ in dependence on the temperature $T$ for various values of magnetic field $h$
[symbols: 
spin system with $J_2=4$,
lines:
lattice-gas model].
Note that the short-dashed ($h=4.5$) and the double-dashed line ($h=5.5$) coincide 
because of the particle-hole symmetry inherent in the lattice-gas model. 
For the spin model the corresponding symbols (up-triangles and circles) also coincide at lower temperatures.}
\label{fig5}
\end{center}
\end{figure}
We observe a very good agreement until $T=0.5$, 
whereas the hard-dimer description is not appropriate at that temperature.  
For $T=1$  deviations between exact diagonalization data and lattice-gas predictions become noticeable.
However, increasing of $J_2$  to $J_2=10$ 
the exact diagonalization result is again indistinguishable from the lattice-gas predictions
(see dotted line and pentagons in the upper panel of Fig.~\ref{fig5}).

It is important to note,
that the obtained results for the thermodynamic quantities refer not only to
finite systems
shown in Fig.~\ref{fig5},
but also to thermodynamically large systems.
The thermodynamic quantities in the limit $N\to\infty$ are given in the
Appendix~\ref{a} by Eqs. (\ref{A.16}) -- (\ref{A.20}).
Hence our findings for the thermodynamics of the frustrated two-leg ladder with $J_2>J_2^c$
together the ground-state analysis given Ref.~\onlinecite{ho_mi_tr} 
lead to a comprehensive description of that frustrated quantum spin model 
in the strong coupling regime.

\subsection{Frustrated bilayer}
\label{sec6b}

Next we consider the lattice-gas model with finite repulsion,
that is relevant for the frustrated bilayer (\ref{2.01}), 
i.e., a lattice-gas of squares on the square lattice, where partial overlap is allowed, 
cf. Sec.~\ref{sec4}.
For small finite lattice-gas systems we use exact formulas for thermodynamic quantities
(see Appendix \ref{b}).
For large finite lattice-gas systems we perform classical Monte Carlo simulations\cite{footnote2}
(see also Appendix \ref{b}).

We start with a brief summary of the known results 
for the phase diagram of the corresponding square-lattice Ising antiferromagnet 
with nearest-neighbor exchange $\cal{J}$ 
in a field $\Gamma$ (\ref{6.07}),\cite{mueller-hartman,wu,wang,betts,viana}
which sets the benchmarks in our further discussion.
In contrast to the one-dimensional case,
the two-dimensional model is known to have an antiferromagnetic long-range order
within a restricted part of the half-plane ``magnetic field $\Gamma$ -- temperature $T$''.
A critical line separating the ordered regime
along which thermodynamic quantities become singular
has been discussed in many papers.\cite{mueller-hartman,wu,wang,betts,viana}
Several closed-form formulas of the critical line $T_c({\Gamma})$
were suggested and compared with numerical results.
Clearly,
along the line $T=0$ the antiferromagnetic phase exists
if $\vert {\Gamma}\vert < 4{\cal{J}}$,
whereas
along the line ${\Gamma}=0$ the antiferromagnetic phase exists below
$T_0/{\cal{J}}=2/\ln(\sqrt{2}+1)\approx 2.269\,185$
(Onsager's zero-field critical point\cite{onsager}).
For the lattice-gas model the corresponding critical line  $T_c(\mu)$ is in the half-plane $\mu$ -- $T$, 
and we get 
$T_0/V=1/[2\ln(\sqrt{2}+1)]\approx 0.567\,296$ 
at $\mu=2V$.
The critical line crosses the $\mu$-axis at $\mu=0$ and $\mu=4V$.
For the initial frustrated quantum Heisenberg bilayer we have to set $V=J_1$. 
Then at $T=0$ the long-range ordered phase occurs if $h_1-4J_1<h<h_1$,
i.e., for magnetic fields within the one-half magnetization plateau $h_2<h<h_1$.
The corresponding critical line  $T_c(h)$ is in the half-plane $h$ -- $T$, 
and we get the maximal critical temperature 
$T_0= T_c[h=(h_1-h_2)/2]=J_1/[2\ln(\sqrt{2}+1)]\approx 0.567\,296 J_1$.

Hence, 
in the frustrated bilayer we have various possibilities 
to pass from the disordered to the long-range ordered phase
(where localized magnons occupied only one of the two sublattices):
(i) Fixing the field $h$, $h_2<h<h_1$,
and decreasing  of temperature $T$ to $T < T_c(h)$.
(ii) Fixing the temperature $T$, $T<J_1/[2\ln(\sqrt{2}+1)]$, 
and decreasing of $h$ starting from above $h_1$.
(iii) Fixing the temperature $T$, $T<J_1/[2\ln(\sqrt{2}+1)]$, 
and increasing of $h$ starting from below $h_2$.
Crossing the critical line $T_c(h)$
the critical behavior is that of the two-dimensional Ising model.
In Fig.~\ref{fig6} we show the phase diagram of the frustrated bilayer
which is a retranslation of the corresponding phase diagram
of the square-lattice Ising antiferromagnet in a field.\cite{mueller-hartman,wu,wang,betts,viana}
Note, however, that we have reproduced this phase diagram 
by our Monte-Carlo simulation of the classical lattice-gas model 
(for more details see below).
\begin{figure}
\begin{center}
\includegraphics[clip=on,width=7.5cm]{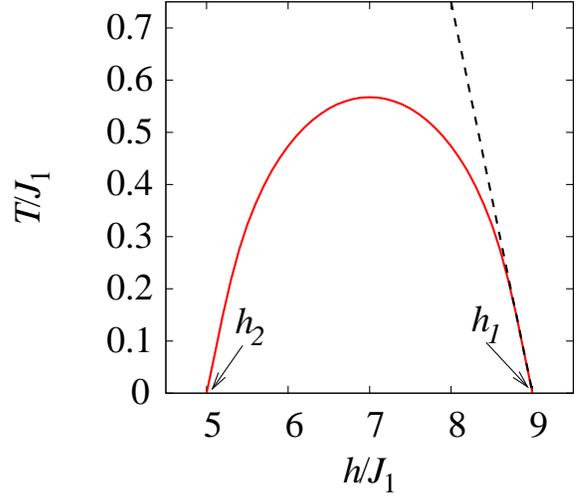}
\end{center}
\caption{(Color online)
Phase diagram of the $s=1/2$ Heisenberg antiferromagnet in a magnetic field 
on the frustrated bilayer lattice with $J_1=1$, $J_2=5$ 
(i.e., $h_1=9$, $h_2=5$)
in the half-plane ``magnetic field $h$ -- temperature $T$''.
The critical line $T_c(h)$ (solid line) separates the long-range ordered
phase, in which the localized  magnons occupy one sublattice and the other one is empty
(below the curve)
and the disordered phase with a random distribution of localized magnons 
(above the curve).
The dashed line corresponds to the critical line according to the hard-square description.}
\label{fig6}
\end{figure}

A short remark about the hard-square case,
which describes relevant physics of the frustrated bilayer (\ref{2.01}) around $h_1$ and small $T$,
is expedient here.
The critical line as it follows from the hard-square model reads:
$T_c(h)=(h_1-h)/\ln z_c$ with $\ln z_c\approx 1.3340$ 
(the dashed line in Fig.~\ref{fig6}),
see Refs.~\onlinecite{baxter_hs,d&r_c}.
The critical behavior 
which emerges while crossing the curve $T_c(h)=(h_1-h)/\ln z_c$ 
for the hard-square model 
also belongs to the two-dimensional Ising model universality class.\cite{baxter_hs,d&r_c}
The results shown in Fig.~\ref{fig6} demonstrate 
that the hard-square phase diagram coincides with the lattice-gas phase diagram only 
around the point $h=h_1$ and $T \lesssim 0.3$.

To estimate the validity of the lattice-gas phase diagram
for the quantum Heisenberg antiferromagnet on the frustrated bilayer lattice (\ref{2.01})
we again compare exact diagonalization data for finite bilayer spin systems
with the lattice-gas predictions,
see Fig.~\ref{fig7}.
It is also useful to compare these results with corresponding ones obtained using the hard-square model,
see Fig.~\ref{fig4}.
\begin{figure}
\begin{center}
\includegraphics[clip=on,width=7.5cm,angle=0]{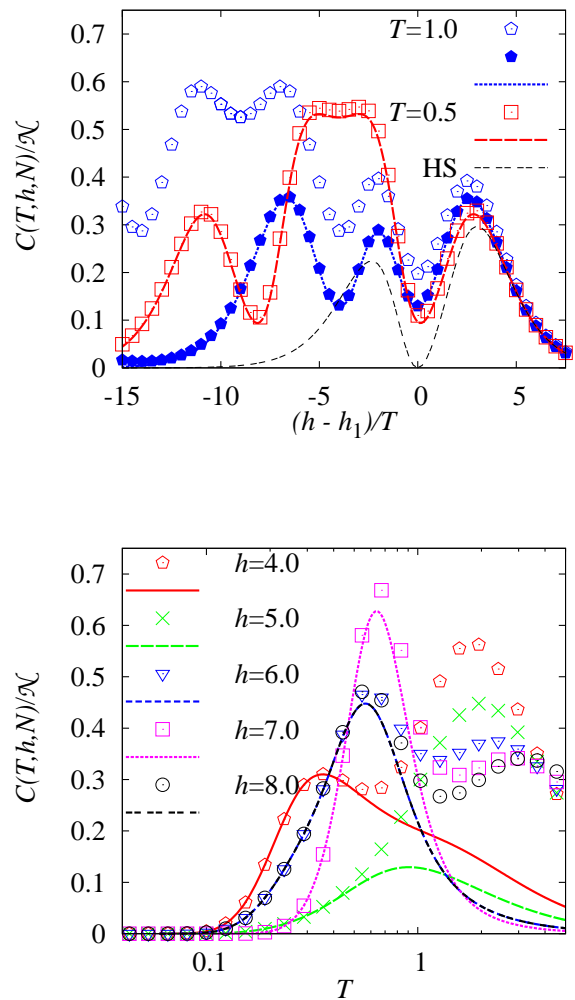}
\caption
{(Color online)
Specific heat $C$ for the frustrated bilayer with $J_1=1$, $J_2=5$ ($h_1=9$, $h_2=5$)
and $J_1=1$, $J_2=10$ ($h_1=14$, $h_2=10$):
Exact diagonalization data versus lattice-gas predictions.
Upper panel:
$C$ in dependence on the hard-core parameter $(h-h_1)/T$ for $N=20$ (${\cal{N}}=10$) and $T=0.5$ and $T=1$
[open symbols:
spin system with $J_2=5$,
filled pentagons:
spin system with $J_2=10$,
long-dashed and dotted lines:
lattice-gas model,
double-dashed line:
hard-square model].
Lower panel:
$C$ in dependence on the temperature $T$ for various values of magnetic field $h$ for $N=16$ (${\cal{N}}=8$)
[symbols:
spin system with $J_2=5$,
lines:
lattice-gas model].
The lines for $h=6$ and $h=8$ coincide because of the particle-hole symmetry of the lattice-gas model.}
\label{fig7}
\end{center}
\end{figure}
From the results reported in Fig.~\ref{fig7} one concludes
that the lattice-gas model provides a very good description 
of the considered finite quantum spin system with $J_1=1$, $J_2=5$ 
in a wide range of magnetic fields at least up to temperatures about $T=0.5$ 
Moreover,
if $J_2$ acquires a large value, $J_2=10$, 
lattice-gas predictions remain very good even at $T=1.0$
(see the upper panel in Fig.~\ref{fig7}).
Hence, 
we have evidence that the phase phase diagram presented in Fig.~\ref{fig6} 
is indeed valid for the $s=1/2$ Heisenberg antiferromagnet in a magnetic field 
on the frustrated bilayer lattice.

Let us discuss two further aspects of the data shown in Fig.~\ref{fig7}.
(i)
Comparing lattice-gas results (long-dashed line in the upper panel) 
with hard-square results (double-dashed line in the upper panel) 
the limited temperature range of validity of the hard-square picture is obvious. 
(ii) 
The extra low-temperature maximum in $C(T)$ is present in a wide range of magnetic fields. 
It is well described by the lattice-gas model.

After having demonstrated the quality of the lattice-gas description 
of the low-temperature thermodynamics for small systems
we consider now thermodynamically large systems.
We have used classical Monte Carlo simulations for the lattice-gas model with finite repulsion 
which reproduce reliably the low-temperature properties of the frustrated bilayer 
in a wide range of magnetic fields.
Note, however, that for the special value of $h=(h_1+h_2)/2=2J_1+J_2$
we face the zero-field square-lattice Ising model
and hence in this limit we have a set of analytical equations for thermodynamic quantities
known from Onsager's solution.\cite{baxter,onsager}
In Fig.~\ref{fig8} 
we show temperature dependences of the specific heat $C$, 
staggered susceptibility $\chi_{{\rm{st}}}$, and the entropy $S$
obtained from classical Monte Carlo simulations.\cite{footnote2}
Due to the particle-hole symmetry 
(which we have confirmed explicitely by our Monte-Carlo calculations) 
the temperature dependence is identical at fields $h=h_2+\Delta h$ and $h=h_1-\Delta h$.
\begin{figure}
\begin{center}
\includegraphics[clip=on,width=7.5cm,angle=0]{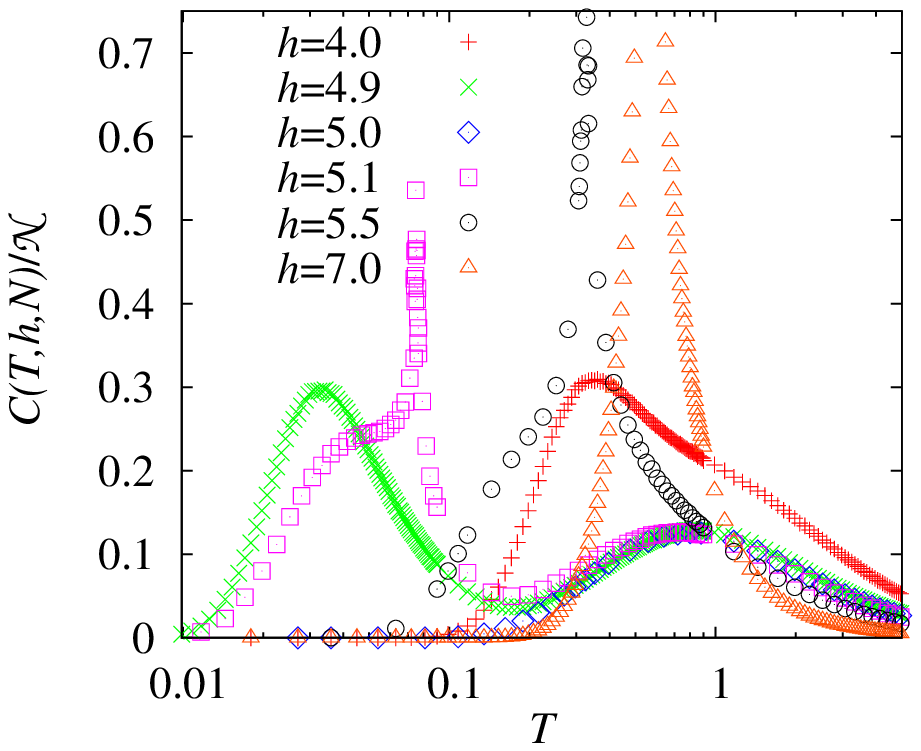}\\
\includegraphics[clip=on,width=7.5cm,angle=0]{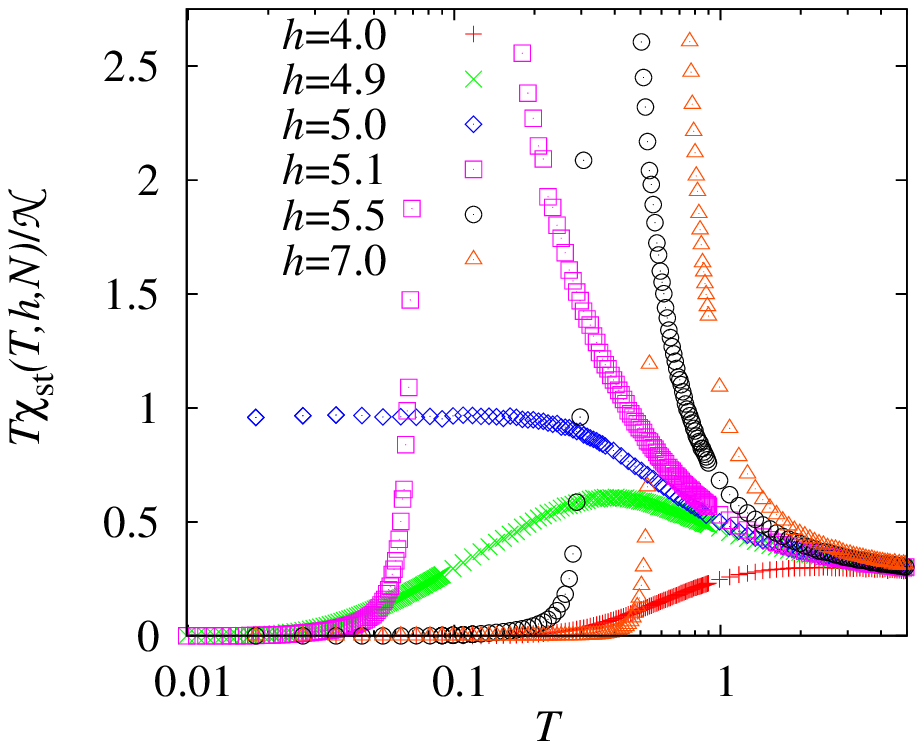}\\
\includegraphics[clip=on,width=7.5cm,angle=0]{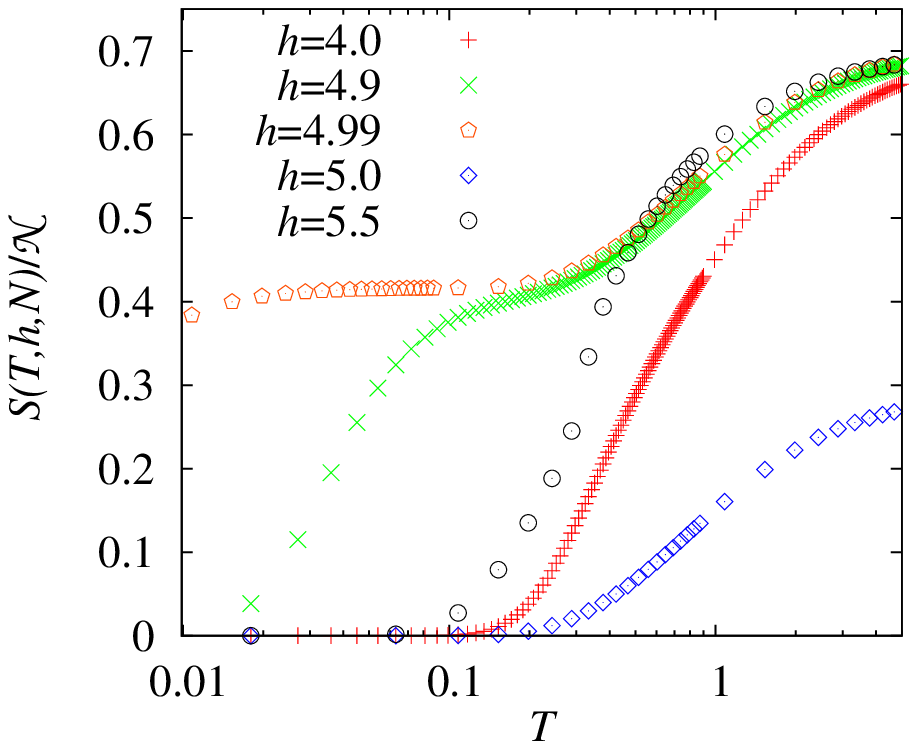}
\caption
{(Color online)
Specific heat (upper panel),
staggered susceptibility (middle panel),
and
entropy (lower panel) versus temperature
for the frustrated bilayer with $J_1=1$ and $J_2=5$ 
($h_1=9$ and $h_2=5$)
for different magnetic fields 
obtained from Monte Carlo simulations for the effective lattice-gas model with finite repulsion $V=J_1$.}
\label{fig8}
\end{center}
\end{figure}
As a main feature, clearly seen for large systems for $h_2 < h < h_1$, 
a divergence of the specific heat $C$ (Fig.~\ref{fig8}, upper panel) 
and the staggered susceptibility $\chi_{{\rm{st}}}$ (Fig.~\ref{fig8}, middle panel) 
appears at a critical temperature 
where the order-disorder phase transition takes place,
cf. the phase diagram shown in Fig.~\ref{fig6}.
As already discussed, 
the critical behavior belongs to the two-dimensional Ising model universality class, 
see Secs.~\ref{sec5b} and \ref{sec6b}. 
Note that the low-temperature maximum in $C(T)$ 
found for smaller systems (see Fig.~\ref{fig7}, lower panel) 
is masked by the logarithmic divergence. 
However, for magnetic fields near $h_2$ and $h_1$ a characteristic bump below the divergence occurs.
The high-temperature maximum in $C(T)$ present for the spin system 
cannot be described correctly by the lattice-gas model.
Another feature worth to be mentioned 
is the behavior of the staggered susceptibility $\chi_{{\rm{st}}}$ for $T \to 0$. 
While there is an exponential decay of $\chi_{{\rm{st}}}$ to zero for $h_2 < h < h_1$, 
precisely at $h=h_1$ and $h=h_2$ it diverges as $1/T$.
With respect to the temperature dependence of the entropy 
shown in the lower panel of Fig.~\ref{fig8} 
it is worthwhile to make the following remark.
To obtain the entropy $S(T,h,N)$ we perform integration according to Eq. (\ref{B.08}).
Note, however, that in Eq. (\ref{B.08}) the contribution at
$T=0$, i.e., $S(T=0,h,N)$, is not included.
This is correct if $h\ne h_1$ or $h\ne h_2$.
However, for $h=h_1$ or $h=h_2$ there is a nonzero ground-state residual entropy 
which is therefore missed in the corresponding curve (diamonds) in the lower panel of Fig.~\ref{fig8}.
Taking into account the constant of integration 
$S(T=0,h_1,N)=S(T=0,h_2,N)=\ln[\kappa(1)]N/2$
leads to a shift of the curves $S(T,h_1,N)/{\cal{N}}$ and $S(T,h_2,N)/{\cal{N}}$ upward by $\approx 0.4075$
and recovers a monotonic dependence of temperature profiles $S(T)$ as $h$ varies from $4.0$ to $5.5$,
see Fig.~\ref{fig8}, lower panel.
However, if the magnetic field is close to $h_1$ or $h_2$ 
(see crosses for $h=4.9$ and pentagons for $h=4.99$ in Fig.~\ref{fig8}) 
as a remnant of the residual entropy present for  $h=h_1$ and
$h=h_2$ the entropy remains large up to quite low temperatures $T \sim
|h-h_2|$ or 
$|h-h_1|$, respectively.

\section{Conclusions}
\label{sec7}

In the present paper we have demonstrated 
that the thermodynamic quantities of two particular quantum many-body systems, 
namely
the frustrated ladder and bilayer $s=1/2$ Heisenberg antiferromagnets in a magnetic field $h$ (\ref{2.01}),
can be obtained via classical lattice-gas-model calculations in a wide range of magnetic fields.
For the one-dimensional ladder model 
by means of the  transfer-matrix method even a complete analytical description is possible, 
whereas for
the two-dimensional bilayer model 
well elaborated classical Monte Carlo simulations can be used.
The reason for this significant simplification 
lies in the  simple structure of low-energy levels of the quantum spin system
which emerges due to frustrations in the strong-coupling regime.
The ground-state magnetization curve in this regime exhibits plateaus 
at zero magnetization and at one-half of the saturation magnetization. 
The classical lattice-gas model leads to an excellent description of the quantum spin models 
up to quite large temperatures of the order of the exchange constants   
in the field region of the one-half plateau, 
i.e., for $h_2 < h < h_1$, 
as well as magnetic fields slightly below $h_2$ and above $h_1$.

Some prominent features of the considered systems are as follows:
a ground-state residual entropy at $h= h_2$  and $ h = h_1$
that may be of particular interest for magnetic cooling\cite{zhi1,zhi2,zhi3,zhi4,d&r_a,d&r_b,schnack}
and 
a second order phase transition at a critical temperature $T_c(h) > 0$, $h_2 < h < h_1$ 
found for the two-dimensional bilayer system, 
where the critical behavior is that of the two-dimensional Ising antiferromagnet.

Finally, we mention that lattice-gas approach elaborated in the present paper 
can be extended to similar (although different) models, 
such as the frustrated three-leg ladder\cite{ho_mi_tr} in one dimension
or 
the bilayer systems consisting of two triangular or honeycomb lattices.
We leave the discussion of these models for further studies.

\section*{Acknowledgments}

The numerical calculations were performed using J.~Schulenburg's {\it{spinpack}}.
The authors thank A.~Honecker and N.~Ivanov for discussions.
The present study was supported by the DFG (projects Ri615/18-1 and Ri615/19-1).
O.~D. acknowledges the kind hospitality of the Magdeburg University in 2009 and 2010
and of the MPIPKS-Dresden in 2010 during the International Workshop on Perspectives in Highly Frustrated Magnetism.

\appendix
\section{One-dimensional lattice-gas models}
\label{a}

In this appendix
we collect some analytical results
for the one-dimensional 
(i) hard-dimer model 
and 
(ii) lattice-gas model with finite nearest-neighbor repulsion
obtained by means of the transfer-matrix method. 
These formulas can be used to calculate the relevant physical quantities at low temperatures 
for the $s=1/2$ frustrated Heisenberg two-leg ladder.

We start with the one-dimensional hard-dimer model,
see Eqs. (\ref{5.04}) and (\ref{5.05}).
For the entropy, the specific heat, the average number of hard dimers, the uniform susceptibility,
and the staggered susceptibility after simple but lengthy calculations we get
\begin{widetext}
\begin{eqnarray}
\frac{S(T,h,N)}{{\cal{N}}}
=\frac{1}{{\cal{N}}}\ln\left({\lambda_+}^{\cal{N}}+{\lambda_-}^{\cal{N}}\right)
-\frac{(\ln z)z}{\sqrt{1+4z}}
\frac{{\lambda_+}^{{\cal{N}}-1}-{\lambda_-}^{{\cal{N}}-1}}{{\lambda_+}^{\cal{N}}+{\lambda_-}^{\cal{N}}},
\label{A.01}
\end{eqnarray}
\begin{eqnarray}
\frac{C(T,h,N)}{{\cal{N}}}
=\frac{(\ln z)^2z}{\sqrt{1+4z}}\left(1-\frac{2z}{1+4z}\right)
\frac{{\lambda_+}^{{\cal{N}}-1}-{\lambda_-}^{{\cal{N}}-1}}{{\lambda_+}^{\cal{N}}+{\lambda_-}^{\cal{N}}}
\nonumber\\
+
\left[\frac{(\ln z)z}{\sqrt{1+4z}}\right]^2
\left[
({\cal{N}}-1)
\frac{{\lambda_+}^{{\cal{N}}-2}+{\lambda_-}^{{\cal{N}}-2}}{{\lambda_+}^{\cal{N}}+{\lambda_-}^{\cal{N}}}
-
{\cal{N}}
\left(\frac{{\lambda_+}^{{\cal{N}}-1}-{\lambda_-}^{{\cal{N}}-1}}{{\lambda_+}^{\cal{N}}+{\lambda_-}^{\cal{N}}}\right)^2
\right],
\label{A.02}
\end{eqnarray}
\begin{eqnarray}
\frac{M(T,h,N)}{{\cal{N}}}=1-\frac{\overline{n}}{{\cal{N}}},
\;\;\;
\frac{\overline{n}}{{\cal{N}}}
=\frac{z}{\sqrt{1+4z}}\frac{{\lambda_+}^{{\cal{N}}-1}-{\lambda_-}^{{\cal{N}}-1}}{{\lambda_+}^{\cal{N}}+{\lambda_-}^{\cal{N}}},
\label{A.03}
\end{eqnarray}
\begin{eqnarray}
\frac{T\chi(T,h,N)}{{\cal{N}}}
=\frac{z}{\sqrt{1+4z}}\left(1-\frac{2z}{1+4z}\right)
\frac{{\lambda_+}^{{\cal{N}}-1}-{\lambda_-}^{{\cal{N}}-1}}{{\lambda_+}^{\cal{N}}+{\lambda_-}^{\cal{N}}}
\nonumber\\
+
\left(\frac{z}{\sqrt{1+4z}}\right)^2
\left[
({\cal{N}}-1)
\frac{{\lambda_+}^{{\cal{N}}-2}+{\lambda_-}^{{\cal{N}}-2}}{{\lambda_+}^{\cal{N}}+{\lambda_-}^{\cal{N}}}
-
{\cal{N}}
\left(\frac{{\lambda_+}^{{\cal{N}}-1}-{\lambda_-}^{{\cal{N}}-1}}{{\lambda_+}^{\cal{N}}+{\lambda_-}^{\cal{N}}}\right)^2
\right],
\label{A.04}
\end{eqnarray}
\begin{eqnarray}
\frac{T\chi_{{\rm{st}}}(T,h,N)}{{\cal{N}}}
=
\frac{z}{2}
\frac{{\lambda_+}^{{\cal{N}}-1}+{\lambda_-}^{{\cal{N}}-1}}{{\lambda_+}^{{\cal{N}}}+{\lambda_-}^{{\cal{N}}}}
+
\frac{z}{2\sqrt{1+4z}}
\frac{{\lambda_+}^{{\cal{N}}-1}-{\lambda_-}^{{\cal{N}}-1}}{{\lambda_+}^{{\cal{N}}}+{\lambda_-}^{{\cal{N}}}},
\label{A.05}
\end{eqnarray}
\end{widetext}
respectively.
We recall that here $z=e^{(h_1-h)/T}$, $\ln z=(h_1-h)/T$,
and $\lambda_{\pm}=(1\pm\sqrt{1+4z})/2$ [see Eq. (\ref{5.04})].

In the limit ${\cal{N}}\to\infty$ the formulas (\ref{A.01}) -- (\ref{A.05}) become much simpler
\begin{eqnarray}
\frac{S(T,h,N)}{{\cal{N}}}
=\ln{\lambda_+}
-\frac{(\ln z)z}{\sqrt{1+4z}}
\frac{1}{{\lambda_+}},
\label{A.06}
\end{eqnarray}
\begin{eqnarray}
\frac{C(T,h,N)}{{\cal{N}}}
=\frac{(\ln z)^2z}{\left(1+4z\right)^{\frac{3}{2}}},
\label{A.07}
\end{eqnarray}
\begin{eqnarray}
\frac{M(T,h,N)}{{\cal{N}}}=1-\frac{\overline{n}}{{\cal{N}}},
\;\;\;
\frac{\overline{n}}{{\cal{N}}}
=\frac{z}{\sqrt{1+4z}}\frac{1}{\lambda_+},
\label{A.08}
\end{eqnarray}
\begin{eqnarray}
\frac{T\chi(T,h,N)}{{\cal{N}}}
=\frac{z}{\left(1+4z\right)^{\frac{3}{2}}},
\label{A.09}
\end{eqnarray}
\begin{eqnarray}
\frac{T\chi_{{\rm{st}}}(T,h,N)}{{\cal{N}}}
=
\frac{z}{\sqrt{1+4z}}.
\label{A.10}
\end{eqnarray}

We turn to the one-dimensional lattice-gas model with finite repulsion,
see Eqs. (\ref{6.08}) and (\ref{6.09}).
For the entropy, the specific heat, the average number of hard dimers, the uniform susceptibility,
and the staggered susceptibility after simple but lengthy calculations we get
\begin{widetext}
\begin{eqnarray}
\frac{S(T,h,N)}{{\cal{N}}}
=\frac{1}{{\cal{N}}}\ln\left({\lambda_+}^{\cal{N}}+{\lambda_-}^{\cal{N}}\right)
-\frac{2(\ln z)z-(\ln w)w(1-w)}{2\sqrt{(1-w)^2+4z}}
\frac{{\lambda_+}^{{\cal{N}}-1}-{\lambda_-}^{{\cal{N}}-1}}{{\lambda_+}^{\cal{N}}+{\lambda_-}^{\cal{N}}}
-\frac{1}{2}(\ln w)w
\frac{{\lambda_+}^{{\cal{N}}-1}+{\lambda_-}^{{\cal{N}}-1}}{{\lambda_+}^{\cal{N}}+{\lambda_-}^{\cal{N}}},
\label{A.11}
\end{eqnarray}
\begin{eqnarray}
\frac{C(T,h,N)}{{\cal{N}}}
=
\frac{b_+{\lambda_+}^{{\cal{N}}-1}+b_-{\lambda_-}^{{\cal{N}}-1}}{{\lambda_+}^{\cal{N}}+{\lambda_-}^{\cal{N}}}
+
({\cal{N}}-1)
\frac{a_+^2{\lambda_+}^{{\cal{N}}-2}+a_-^2{\lambda_-}^{{\cal{N}}-2}}{{\lambda_+}^{\cal{N}}+{\lambda_-}^{\cal{N}}}
-
{\cal{N}}
\left(\frac{a_+{\lambda_+}^{{\cal{N}}-1}+a_-{\lambda_-}^{{\cal{N}}-1}}{{\lambda_+}^{\cal{N}}+{\lambda_-}^{\cal{N}}}\right)^2,
\nonumber\\
a_{\pm}=\mp\frac{2(\ln z)z-(\ln w)w(1-w)}{2\sqrt{(1-w)^2+4z}}-\frac{1}{2}(\ln w)w,
\nonumber\\
b_{\pm}=\pm\frac{2(\ln z)^2z-(\ln w)^2w(1-2w)}{2\sqrt{(1-w)^2+4z}}
\mp \frac{\left[2(\ln z)z-(\ln w)w(1-w)\right]^2}{2\left[(1-w)^2+4z\right]^{\frac{3}{2}}}
+\frac{1}{2}(\ln w)^2w,
\label{A.12}
\end{eqnarray}
\begin{eqnarray}
\frac{M(T,h,N)}{{\cal{N}}}=1-\frac{\overline{n}}{{\cal{N}}},
\;\;\;
\frac{\overline{n}}{{\cal{N}}}
=\frac{2z-w(1-w)}{2\sqrt{(1-w)^2+4z}}
\frac{{\lambda_+}^{{\cal{N}}-1}-{\lambda_-}^{{\cal{N}}-1}}{{\lambda_+}^{\cal{N}}+{\lambda_-}^{\cal{N}}}
+\frac{w}{2}
\frac{{\lambda_+}^{{\cal{N}}-1}+{\lambda_-}^{{\cal{N}}-1}}{{\lambda_+}^{\cal{N}}+{\lambda_-}^{\cal{N}}},
\label{A.13}
\end{eqnarray}
\begin{eqnarray}
\frac{T\chi(T,h,N)}{{\cal{N}}}
=
\frac{d_+{\lambda_+}^{{\cal{N}}-1}+d_-{\lambda_-}^{{\cal{N}}-1}}{{\lambda_+}^{{\cal{N}}}+{\lambda_-}^{{\cal{N}}}}
+
({\cal{N}}-1)
\frac{c_+^2{\lambda_+}^{{\cal{N}}-2}+c_-^2{\lambda_-}^{{\cal{N}}-2}}{{\lambda_+}^{\cal{N}}+{\lambda_-}^{\cal{N}}}
-
{\cal{N}}
\left(\frac{c_+{\lambda_+}^{{\cal{N}}-1}+c_-{\lambda_-}^{{\cal{N}}-1}}{{\lambda_+}^{\cal{N}}+{\lambda_-}^{\cal{N}}}\right)^2,
\nonumber\\
c_{\pm}=\pm\frac{2z-w(1-w)}{2\sqrt{(1-w)^2+4z}}+\frac{w}{2},
\nonumber\\
d_{\pm}=
\pm\frac{2z-w(1-2w)}{2\sqrt{(1-w)^2+4z}}
\mp \frac{\left[2z-w(1-w)\right]^2}{2\left[(1-w)^2+4z\right]^{\frac{3}{2}}}
+\frac{w}{2},
\label{A.14}
\end{eqnarray}
\begin{eqnarray}
\frac{T\chi_{{\rm{st}}}(T,h,N)}{{\cal{N}}}
=
\frac{z}{2(1+w)}
\frac{{\lambda_+}^{{\cal{N}}-1}+{\lambda_-}^{{\cal{N}}-1}}{{\lambda_+}^{{\cal{N}}}+{\lambda_-}^{{\cal{N}}}}
+
\frac{z}{2\sqrt{(1-w)^2+4z}}
\frac{{\lambda_+}^{{\cal{N}}-1}-{\lambda_-}^{{\cal{N}}-1}}{{\lambda_+}^{{\cal{N}}}+{\lambda_-}^{{\cal{N}}}},
\label{A.15}
\end{eqnarray}
\end{widetext}
respectively.
We recall that here $z=e^{(h_1-h)/T}$, $\ln z=(h_1-h)/T$,
$\lambda_{\pm}=[1+w\pm\sqrt{(1-w)^2+4z}]/2$ [see Eq. (\ref{6.08})]
and we have also introduced the notations $w=e^{(h_1-h-J_1)/T}$, $\ln w=(h_1-h-J_1)/T$.
Evidently in the limit $w\to 0$ Eqs. (\ref{A.11}) -- (\ref{A.15}) transform into Eqs. (\ref{A.01}) -- (\ref{A.05}).

In the limit ${\cal{N}}\to\infty$ the formulas (\ref{A.11}) -- (\ref{A.15}) become much simpler
\begin{eqnarray}
\frac{S(T,h,N)}{{\cal{N}}}
=\ln{\lambda_+}
\nonumber\\
-
\left[\frac{2(\ln z)z-(\ln w)w(1-w)}{2\sqrt{(1-w)^2+4z}}
+\frac{1}{2}(\ln w)w\right]
\frac{1}{{\lambda_+}},
\label{A.16}
\end{eqnarray}
\begin{eqnarray}
\frac{C(T,h,N)}{{\cal{N}}}
=\frac{b_+}{\lambda_+}-\frac{a_+^2}{\lambda_+^2},
\label{A.17}
\end{eqnarray}
\begin{eqnarray}
\frac{M(T,h,N)}{{\cal{N}}}=1-\frac{\overline{n}}{{\cal{N}}},
\nonumber\\
\frac{\overline{n}}{{\cal{N}}}
=\left[\frac{2z-w(1-w)}{2\sqrt{(1-w)^2+4z}}+\frac{w}{2}\right]\frac{1}{\lambda_+},
\label{A.18}
\end{eqnarray}
\begin{eqnarray}
\frac{T\chi(T,h,N)}{{\cal{N}}}
=\frac{d_+}{\lambda_+}-\frac{c_+^2}{\lambda_+^2},
\label{A.19}
\end{eqnarray}
\begin{eqnarray}
\frac{T\chi_{{\rm{st}}}(T,h,N)}{{\cal{N}}}
=
\frac{z}{(1+w)\sqrt{(1-w)^2+4z}}.
\label{A.20}
\end{eqnarray}
Again after inserting $w=0$ into Eqs. (\ref{A.16}) -- (\ref{A.20}) we obtain Eqs. (\ref{A.06}) -- (\ref{A.10})
as it should be.

\section{Two-dimensional lattice-gas models}
\label{b}

In this appendix
we collect some formulas
for the two-dimensional 
(i) hard-square model 
and 
(ii) lattice-gas model with finite nearest-neighbor repulsion
which we use in our direct calculations of thermodynamic quantities for small systems
and classical Monte Carlo simulations for large systems.

We can obtain thermodynamic quantities for finite hard-square models by direct calculations
starting from the definition of the grand-canonical partition function
\begin{eqnarray}
\Xi_{\rm{hc}}(z,{\cal{N}})=\sum_{n=0}^{\frac{{\cal{N}}}{2}}{\cal{Z}}_{\rm{hc}}(n,{\cal{N}})z^n
\label{B.01}
\end{eqnarray}
and knowing the canonical partition functions ${\cal{Z}}_{\rm{hc}}(n,{\cal{N}})$ for $n=0,1,\ldots,{\cal{N}}/2$
(calculation of these numbers are feasible for small ${\cal{N}}$).
Really, $\Xi_{\rm{hc}}(z,{\cal{N}})$ is a polynomial of order ${\cal{N}}/2$
and calculations of thermodynamic quantities are doable although rather tedious.
Thus,
for the entropy, the specific heat, the average number of hard squares, and the uniform susceptibility we find
\begin{eqnarray}
S(T,h,N)
=
\ln\left[\Xi_{\rm{hc}}(z,{\cal{N}})\right]
-
(\ln z)\overline{n},
\label{B.02}
\end{eqnarray}
\begin{eqnarray}
C(T,h,N)
=(\ln z)^2\left(\overline{n^2}-{\overline{n}}^2\right),
\label{B.03}
\end{eqnarray}
\begin{eqnarray}
M(T,h,N)
=\frac{N}{2}-\overline{n},
\label{B.04}
\end{eqnarray}
\begin{eqnarray}
T\chi(T,h,N)
=\overline{n^2}-{\overline{n}}^2,
\label{B.05}
\end{eqnarray}
where $z=e^{(h_1-h)/T}$ is the activity
and
$\overline{(\ldots)}=[\sum_{n=0}^{{\cal{N}}/2}{\cal{Z}}_{\rm{hc}}(n,{\cal{N}})z^n(\ldots)]/\Xi_{\rm{hc}}(z,{\cal{N}})$
denotes the grand-canonical average for the considered finite hard-square model of ${\cal{N}}$ sites.

For completeness we give here canonical partition functions for some finite lattices:
${\cal{Z}}_{{\rm{hc}}}(n,8)=1,8,12,8,2$ for $n=0,1,2,3,4$,
${\cal{Z}}_{{\rm{hc}}}(n,10)=1,10,25,20,10,2$ for $n=0,1,2,3,4,5$,
${\cal{Z}}_{{\rm{hc}}}(n,16)=1,16,88,208,228,128,56,16,2$ for $n=0,1,2,3,4,5,6,7,8$.

To calculate the staggered magnetization
\begin{eqnarray}
M_{\rm{st}}(T,h,N)=\overline{n_A}-\overline{n_B}
\label{B.06}
\end{eqnarray}
and the staggered susceptibility
\begin{eqnarray}
T\chi_{\rm{st}}(T,h,N)
=\overline{n_A^2}-{\overline{n_A}}^2+\overline{n_B^2}-{\overline{n_B}}^2
\nonumber\\
-2(\overline{n_An_B}-\overline{n_A}\,\overline{n_B})
\label{B.07}
\end{eqnarray}
we have to introduce the quantities ${\cal{Z}}_{{\rm{hc}}}(n_A,n_B;{\cal{N}})$
which are the numbers of spatial configurations of $n$ hard squares, where
$n_A$ of them occupy the sublattice $A$
and
$n_B=n-n_A$ of them occupy the sublattice $B$.
Obviously,
${\cal{Z}}_{{\rm{hc}}}(n,{\cal{N}})=\sum_{n_A=0}^n{\cal{Z}}_{{\rm{hc}}}(n_A,n_B;{\cal{N}})$.
We have to refine the definition of the grand-canonical average
making it sensitive to the sublattice indices.
When the staggered component of the activity vanishes,
i.e., $z_A=z_B=z$,
we have
$\overline{(\ldots)}
=
[\sum_{n=0}^{{\cal{N}}/2} z^n \sum_{n_A=0}^n
{\cal{Z}}_{\rm{hc}}(n_A,n_B;{\cal{N}})(\ldots)]/\Xi_{\rm{hc}}(z,{\cal{N}})$.

We give here ${\cal{Z}}_{{\rm{hc}}}(n_A,n_B;{\cal{N}})$ for some finite lattices:
${\cal{Z}}_{{\rm{hc}}}(1,0;8)=4$;
${\cal{Z}}_{{\rm{hc}}}(2,0;8)=6$, ${\cal{Z}}_{{\rm{hc}}}(1,1;8)=0$;
${\cal{Z}}_{{\rm{hc}}}(3,0;8)=4$, ${\cal{Z}}_{{\rm{hc}}}(2,1;8)=0$;
${\cal{Z}}_{{\rm{hc}}}(4,0;8)=1$, ${\cal{Z}}_{{\rm{hc}}}(3,1;8)={\cal{Z}}_{{\rm{hc}}}(2,2;8)=0$
for ${\cal{N}}=8$,
${\cal{Z}}_{{\rm{hc}}}(1,0;10)=5$;
${\cal{Z}}_{{\rm{hc}}}(2,0;10)=10$, ${\cal{Z}}_{{\rm{hc}}}(1,1;10)=5$;
${\cal{Z}}_{{\rm{hc}}}(3,0;10)=10$, ${\cal{Z}}_{{\rm{hc}}}(2,1;10)=0$;
${\cal{Z}}_{{\rm{hc}}}(4,0;10)=5$, ${\cal{Z}}_{{\rm{hc}}}(3,1;10)={\cal{Z}}_{{\rm{hc}}}(2,2;10)=0$,
${\cal{Z}}_{{\rm{hc}}}(5,0;10)=1$, ${\cal{Z}}_{{\rm{hc}}}(4,1;10)={\cal{Z}}_{{\rm{hc}}}(3,2;10)=0$
for ${\cal{N}}=10$.

Formulas (\ref{B.03}) -- (\ref{B.07}) are also used for obtaining Monte Carlo predictions for large hard-square systems.
In Monte Carlo simulations we calculate
$\overline{n_A}$, $\overline{n_B}$, $\overline{n^2_A}$, $\overline{n^2_B}$, and $\overline{n_An_B}$
for a given $z=e^{(h_1-h)/T}$.
As a result,
we obtain $\overline{n}=\overline{n_A}+\overline{n_B}$
and hence $M(T,h,N)$ (\ref{B.04})
and
$\overline{n^2}-{\overline{n}}^2
=\overline{n_A^2}-{\overline{n_A}}^2+\overline{n_B^2}-{\overline{n_B}}^2
+2(\overline{n_An_B}-\overline{n_A}\,\overline{n_B})$
and hence $\chi(T,h,N)$ (\ref{B.05}) and $C(T,h,N)$ (\ref{B.03}).
Then the entropy is obtained by integration
\begin{eqnarray}
S(T,h,N)=\int_0^{T}dT^{\prime}\frac{C(T^{\prime},h,N)}{T^{\prime}}.
\label{B.08}
\end{eqnarray}
Moreover,
Monte Carlo data yield $\vert M_{\rm{st}}(T,h,N)\vert$ (\ref{B.06})
and $\chi_{\rm{st}}(T,h,N)$ (\ref{B.07}).

We turn to the two-dimensional lattice with finite nearest-neighbor repulsion.
Starting from the formula for the grand-canonical partition function $\Xi_{\rm{lg}}(T,\mu,{\cal{N}})$ (\ref{6.03})
and the definition of the grand-canonical average
\begin{eqnarray}
\overline{(\ldots)}
=\frac{\sum_{n_1=0,1}\ldots\sum_{n_{\cal{N}}=0,1}e^{-\frac{{\cal{H}}(\{n_m\})}{T}}(\ldots)}{\Xi_{\rm{lg}}(T,\mu,{\cal{N}})}
\label{B.09}
\end{eqnarray}
we immediately get
\begin{eqnarray}
S(T,h,N)
=
\ln\Xi_{\rm{lg}}(T,\mu,{\cal{N}})
+\frac{\overline{{\cal{H}}(\{n_m\})}}{T},
\label{B.10}
\end{eqnarray}
\begin{eqnarray}
C(T,h,N)
=
\frac{\overline{{\cal{H}}^2(\{n_m\})}-\overline{{\cal{H}}(\{n_m\})}^2}{T^2}
\label{B.11}
\end{eqnarray}
for the entropy and the specific heat, respectively.
For the uniform magnetization, the uniform susceptibility,
the staggered magnetization, and the staggered susceptibility
we formally have the same expressions as in Eqs. (\ref{B.04}), (\ref{B.05}), (\ref{B.06}), and (\ref{B.07}),
however, with the grand-canonical average defined in Eq. (\ref{B.09}).
In the limit $V/T\to\infty$ Eqs. (\ref{B.10}), (\ref{B.11}) transform into Eqs. (\ref{B.02}), (\ref{B.03})
since Eq. (\ref{B.09}) becomes the grand-canonical average for the hard-square model
and
$\overline{{\cal{H}}(\{n_m\})}\to-\mu\overline{n}$,
where $\overline{n}$ is the grand-canonical average number of hard squares.

We use formulas (\ref{B.10}), (\ref{B.11}), (\ref{B.04}), (\ref{B.05}), (\ref{B.06}), and (\ref{B.07})
with the grand-canonical average (\ref{B.09})
for direct calculations for small finite  systems
encoding easily the required computations for ${\cal{N}}=8,\,10$ in a short Fortran program.
For large systems we obtain from Monte Carlo simulations
$\overline{{\cal{H}}(\{n_m\})}$ and $\overline{{\cal{H}}^2(\{n_m\})}$
yielding the specific heat $C(T,h,N)$ (\ref{B.11})
and by integration the entropy $S(T,h,N)$, see Eq. (\ref{B.08}).
Furthermore,
we also compute $\overline{n_A}$, $\overline{n_B}$, $\overline{n_A^2}$, $\overline{n_B^2}$, and $\overline{n_An_B}$
to obtain the magnetizations and the susceptibilities, see Eqs. (\ref{B.04}), (\ref{B.05}), (\ref{B.06}), and (\ref{B.07}).


\begin{thebibliography}{99}
\bibitem{footnote1}
The underlying lattice in these cases coincides with an auxiliary lattice
(on which hard-core objects are placed)
which is usually introduced to count the independent localized-magnon state degeneracy,
see Refs.~\onlinecite{zhi1,zhi3,zhi4,d&r_a,d&r_b,d&r_c,d&r_d}.
\bibitem{footnote2}
In our classical Monte Carlo simulations 
of the hard-square model and the lattice-gas model with finite repulsion 
we consider systems of $50 \times 50$, \ldots, $400 \times 400$ sites
(but sometimes even of $800 \times 800$ sites),
exploit the usual Metropolis algorithm,
and perform up to 3 000 000 Monte Carlo steps.
\end{thebibliography}
\end{document}